\documentclass[reprint,aip,jcp,amssymb,amsmath,floatfix]{revtex4-1}
\usepackage{amsmath}
\usepackage{dcolumn}% Align table columns on decimal point
\usepackage{bm}%
\usepackage{graphicx}
\usepackage{natbib}
\usepackage[colorlinks=true,linkcolor=blue]{hyperref}%
\usepackage{algorithm}
\usepackage{algorithmic}
\usepackage{siunitx}

\newcommand{\AVE}[1]{\ensuremath{\left\langle {#1} \right\rangle}}

\newcommand{\NORMtwo}[1]{\ensuremath{\left\lVert {#1} \right\rVert_2}}
\newcommand{\CP}[2]{\ensuremath{0\leq {#1} \perp {#2} \geq 0}}

\newcommand{\bA}{\ensuremath{\bm{A}}}

\newcommand{\bD}{\ensuremath{\bm{D}}}

\newcommand{\bF}{\ensuremath{\bm{F}}}
\newcommand{\bG}{\ensuremath{\bm{G}}}
\newcommand{\bH}{\ensuremath{\bm{H}}}
\newcommand{\bI}{\ensuremath{\bm{I}}}
\newcommand{\bJ}{\ensuremath{\bm{J}}}
\newcommand{\bK}{\ensuremath{\bm{K}}}

\newcommand{\bM}{\ensuremath{\bm{M}}}
\newcommand{\bN}{\ensuremath{\bm{N}}}

\newcommand{\bP}{\ensuremath{\bm{P}}}
\newcommand{\bQ}{\ensuremath{\bm{Q}}}

\newcommand{\bT}{\ensuremath{\bm{T}}}
\newcommand{\bU}{\ensuremath{\bm{U}}}

\newcommand{\bb}{\ensuremath{\bm{b}}}

\newcommand{\bff}{\ensuremath{\bm{f}}}
\newcommand{\bg}{\ensuremath{\bm{g}}}

\newcommand{\bn}{\ensuremath{\bm{n}}}

\newcommand{\bp}{\ensuremath{\bm{p}}}
\newcommand{\bq}{\ensuremath{\bm{q}}}
\newcommand{\br}{\ensuremath{\bm{r}}}
\newcommand{\bs}{\ensuremath{\bm{s}}}

\newcommand{\bx}{\ensuremath{\bm{x}}}
\newcommand{\by}{\ensuremath{\bm{y}}}

\newcommand{\bPhi}{\ensuremath{\bm{\Phi}}}
\newcommand{\bgamma}{\ensuremath{\bm{\gamma}}}

\newcommand{\bsigma}{\ensuremath{\bm{\sigma}}}
\newcommand{\bSigma}{\ensuremath{\bm{\Sigma}}}

\newcommand{\bOmega}{\ensuremath{\bm{\Omega}}}
\newcommand{\bFcal}{\ensuremath{\bm{\mathcal{F}}}}
\newcommand{\bUcal}{\ensuremath{\bm{\mathcal{U}}}}
\newcommand{\bMcal}{\ensuremath{\bm{\mathcal{M}}}}
\newcommand{\bRcal}{\ensuremath{\bm{\mathcal{R}}}}
\newcommand{\bDcal}{\ensuremath{\bm{\mathcal{D}}}}

\newcommand{\bbR}{\ensuremath{{\mathbb{R}}}}

\begin{document}
	
	\title{
	Computing collision stress in assemblies of active spherocylinders: applications of a fast and generic geometric method
	}
	% \title{The collision stress between aspherical rigid particles: theory, algorithm, and application}
	
	\author{Wen Yan}
	\email[]{wyan@flatironinstitute.org, wenyan4work@gmail.com}
	\affiliation{Center for Computational Biology, Flatiron Institute, Simons Foundation, New York,  NY 10010, USA}
	\affiliation{Courant Institute of Mathematical Sciences, New York University, New York, NY 10012, USA}
	\author{Huan Zhang}
	\affiliation{Shanghai Jiao Tong University, Shanghai 200240, People's Republic of China}
	\affiliation{Courant Institute of Mathematical Sciences, New York University, New York, NY 10012, USA}
	\author{Michael J. Shelley}
	\affiliation{Center for Computational Biology, Flatiron Institute, Simons Foundation, New York,  NY 10010, USA}
	\affiliation{Courant Institute of Mathematical Sciences, New York University, New York, NY 10012, USA}
	
	\date{\today}%
	%\revised{August 2010}%
\begin{abstract}
	In this work we provide a solution to the problem of computing collision stress in particle-tracking simulations.
	First, a formulation for the collision stress between particles is derived as an extension of the virial stress formula to general-shaped particles with uniform or non-uniform density.
	Second, we describe a collision-resolution algorithm based on geometric constraint minimization which eliminates the stiff pairwise potentials in traditional methods.
	The method is validated with a comparison to the equation of state of Brownian spherocylinders.
	Then we demonstrate the application of this method in several emerging problems of soft active matter.
\end{abstract}
	
\maketitle
	
	%\tableofcontents

\section{Introduction}
\label{sec:intro}
% importance of collision stress
Computing bulk collision stress is one of the key statistical tasks in simulations of many particle systems for both underdamped and overdamped, ranging from the molecular to the granular-flow scale.
Collision stress is important because it contributes significantly to the Equation of State (EOS) and rheological properties of such systems.
Notable examples include phase transitions in liquid crystals\cite{bolhuis_tracing_1997} and Active Brownian Particles \cite{TakatoriSwimPressureStress2014}, and the jamming and glassy states of spherical colloids \cite{wang_constant_2015}.

% current weakness 1, geometry
In simulations involving point particles, the collision stress can be computed with the usual virial formula $\AVE{\bx\bF_C}$, where the moment $\bx$ is the vector connecting each pair of point particles, and $\bF_C$ is the collision force between each pair. 
Collision stress in spherical particles of uniform density can be computed in the same way.
A large volume of work can be found in literature discussing all aspects of how to compute collision stress for various systems, but two problems remain.
First, there remains some disagreement about how to compute the stress generated from one pair of colliding asphericalal particles or spherical particles with nonuniform density. 
Some earlier work uses the same virial formula as in the point particle case, where the moment vector $\bx$ is the vector connecting the center-of-mass of two particles\cite{rebertus_molecular_1977}.
In some work for slender rods, the moment vector $\bx$ is taken to be the minimal distance between two center-lines of the colliding pair of rods \cite{SnookNormalstressdifferences2014}.
In work for granular flow involving spherical particles, the virial contribution is integrated over the two particles' volumes, instead of picking only one point on each particle \cite{Campbellstresstensortwodimensional1986,Campbellstresstensorsimple1989}.
To our best knowledge, such different approaches haven't been systematically examined.

% current weakness 2, stiff potential
Another crucial problem is how to detect and resolve the collisions.
Traditionally, collisions are resolved by including a pairwise repulsive force, usually governed by Lennard-Jones (LJ) or Weeks-Chandler-Andersen (WCA) potential, and particle trajectories are integrated over time.
There are two key problems in this traditional approach. 
First, the pairwise repulsive potentials cause stiffness in the time-integrator and require very small time-step sizes.
Second, such pairwise potentials always extend repulsive forces over a finite range, and therefore the collisions are resolved as if the particles were soft and deformable.
For example, in work on Brownian rods \cite{tao_brownian_2005-1} the authors reported an `effective' diameter that is equal to around $90\%$ of the imposed rod diameter, because the repulsive forces cannot be infinitely stiff.
Other collision-resolving methods have been developed upon the idea of geometric constraints.
In these methods the collision forces are not computed using an intermediate repulsive potential.
Instead, the forces are \emph{solved} for by imposing the geometric constraint that at the end of the current time-step, the particles cannot overlap.
The method by \citet{Maurytimesteppingschemeinelastic2006} is one notable example in this style, but his formulation does not preserve the pairwise collision network and therefore the necessary information to compute collision stress is lost.
Another method by \citet{tasora_large-scale_2008} follows similar ideas, but constructs the geometrical constraint problem in a way that the pairwise collision network and Newton's third law are all preserved.
This method has been successfully applied in underdamped granular flow problems.

% outline and organization
In this work we present a complete and efficient solution to resolve collisions and to compute collision stress.
We first resolve the discrepancies in the pairwise contribution to collision stress in Section~\ref{sec:paircolstress}.
The formula is derived as an extension to the virial stress formula in the most general settings, considering the momentum transfer throughout the entire volume of the particles.
We then describe a collision resolution method for overdamped systems in Section~\ref{sec:colalgo}, together with a fast and parallel solver, as a generalization of the method by \citet{tasora_matrix-free_2011}.
In particular, we allow the mobility matrix $\bMcal$ to be computed by any method or approximations which keeps $\bMcal$ symmetric-positive-definite (SPD).
Our method is validated in Section~\ref{sec:benchsylinder} by simulating Brownian spherocylinders and comparing the measured EOS with the classic work by \citet{bolhuis_tracing_1997}.
In Section~\ref{sec:application} we demonstrate the application of our solution by measuring the collision stress in soft active matter systems, including self-propelled rods and growing-dividing cells.

\section{Pairwise collision stress}
\label{sec:paircolstress}
In this section we consider the collision stress generated by one pair of particles in the most general setting, for both underdamped and overdamped systems.
We make only the following assumptions of the collision between two rigid bodies:
\begin{itemize}
	\item The collision force is between one point on particle 1 and one point on particle 2.
	\item The collision process is almost instantaneous.
	\item Newton's third law is satisfied.
\end{itemize}

In particular, \emph{no} assumptions are made for the shape, friction, density, etc., of the two particles.
We shall also see that the existence of other forces like gravity does not change the formulae.
Also, the two points where the collision force is transmitted do not have to be on the two particles' surfaces.

\begin{figure}[h]
	\includegraphics[width=\linewidth]{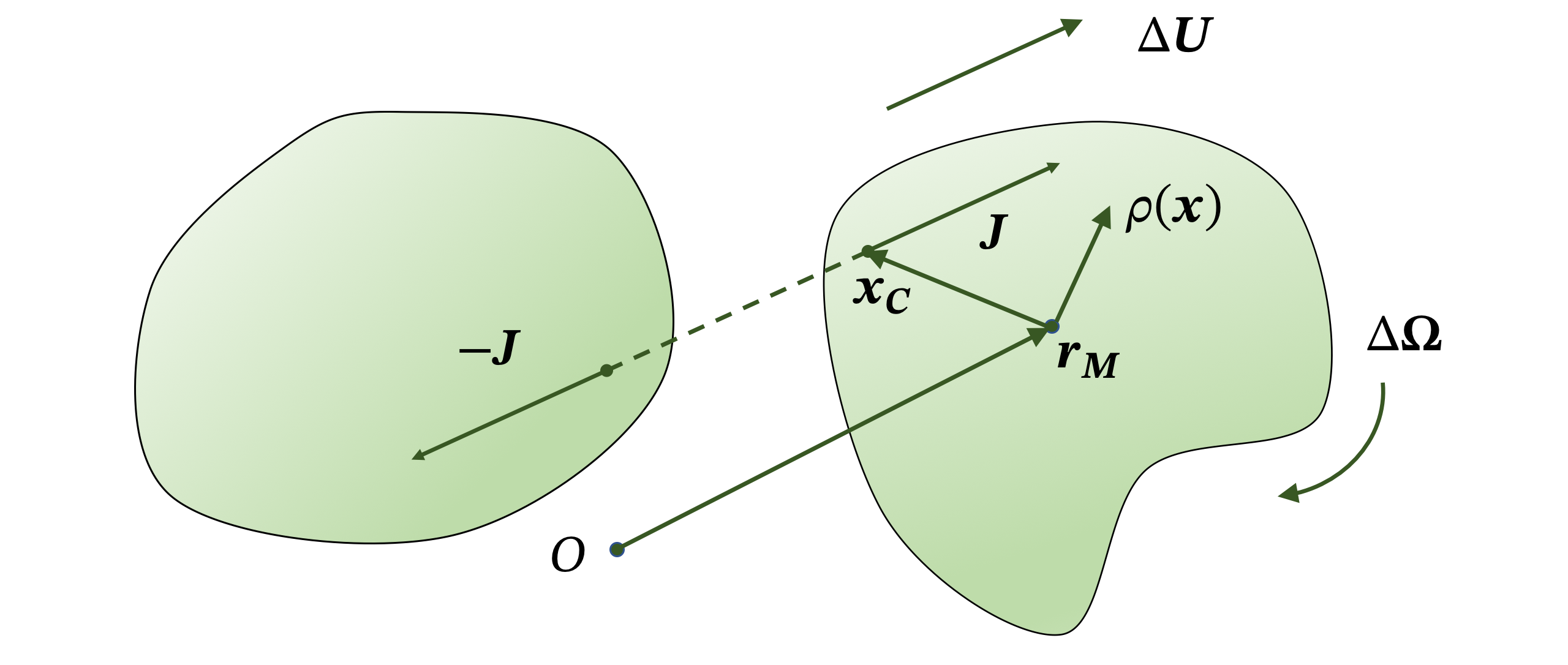}
	\caption{Collision geometry of two arbitrary-shaped rigid bodies. 
	\label{fig:colgeometry} 
	$\br_M$ is the center of mass and $\rho(\bx)$ is the mass density of the particle in the particle frame. 
	$\bx_C$ is the point where the transfer of momentum $\bJ$ happens. 
	$\Delta \bU$ and $\Delta \bOmega$ are the changes in the center of mass velocity and the angular velocity, respectively, due to the action of $\bJ$. }
\end{figure}
We consider the collision geometry shown in Fig.~\ref{fig:colgeometry}.
$O$ is the origin of lab frame, $\br_M$ is the center of mass in the lab frame.
$\bx$ is the location of a mass point relative to the center of mass, and $\bx_C$ is the location of collision in that frame. 
$\bJ$ is the impulse due to this collision event.
For a small duration of collision, $\bJ = \bF_C \delta t$.

\subsection{Governing equations}
Due to symmetry it is sufficient to consider the motion of only one body of the collision pair.
Let the change of velocity and angular velocity due to collision be $\Delta \bU$ and $\Delta \bOmega$.
With Newton's laws we have two equations for translational motion:
\begin{align}
\label{eq:rbeomtrans}
&\int_V \rho\left(\bU + \bOmega \times \bx \right)dV + \bJ \nonumber \\
= &\int_V \rho\left[\left(\bU + \Delta \bU\right) + \left(\bOmega+\Delta\bOmega\right) \times \bx \right]dV,
\end{align}
and rotational motion:
\begin{align}
\label{eq:rbeomrot}
&\int_V \left(\br_M + \bx\right)\times \rho\left(\bU + \bOmega \times \bx \right)dV + \left(\br_M+\bx_C\right)\times\bJ \nonumber \\
= &\int_V \left(\br_M + \bx\right)\times \rho\left[\left(\bU + \Delta \bU\right) + \left(\bOmega+\Delta\bOmega\right) \times \bx \right]dV.
\end{align}

We have the definition of mass $M$ and the moment of inertia tensor $\bG_M$:
\begin{align}
	\int_V \rho dV &= M,\\
	\int_V \rho \left(x^2 \bI - \bx\bx \right) dV &= \bG_M
\end{align}
By definition $\bG_M$ is always symmetric positive definite.
Because $\bx$ is the location in the particle frame relative to the center of mass, we have:
\begin{align}
	\int_V \rho \bx dV &= \bm{0}.
\end{align}
We further define the tensors $\bN$ and $\bQ$ to simplify the tensor notations in the derivation, using
\begin{align}
\bN &=	\int_V \rho \bx \bx dV,\\
\bQ &= \bG_M^{-1}.
\end{align}

Physically, the stress generated by this pair of particles colliding is related to the momentum transfer during the collision, 
which quantitatively, is the integral of the `point-wise virial contribution $\bx\bff\delta t$' over the entire volume of the rigid body, denoted by the tensor $\bs$, for both objects in the collision pair.
In other words, the task is to determine $\bs$ defined as
\begin{align}
	\bs = \int_V \rho(\br_M + \bx) (\Delta \bU + \Delta \bOmega\times \bx) dV,
\end{align}
given the collision force and geometry.
Once $\bs$ is known for both particle 1 and 2, the collision stress generated by this pair is simply:
\begin{align}
\bsigma^{12} = \frac{1}{\delta t}\left(\bs^1 + \bs^2\right).
\end{align}

\subsection{General results}
Equations~(\ref{eq:rbeomtrans}) and~(\ref{eq:rbeomrot}) can be simplified as:
\begin{align}
\bJ &= M \Delta \bU,  \\
\bx_C\times\bJ &= \bG_M \cdot \Delta\bOmega,
\end{align}
where we used the definition of center of mass.
Then $\bs$ can be simplified:
\begin{align}
\bs
&= \br_M \bJ + \int_V \rho \bx \left(\Delta \bU + \Delta \bOmega \times \bx\right) dV.
\end{align}
The first term $\br_M\bJ$ simply corresponds to the virial stress. 
Since $\bU$ is the center-of-mass velocity independent of $\bx$, the integral $\int_V\rho \bx \Delta \bU$ in the second term vanishes by the definition of center of mass. 
We define the integral as $\bs^G$, i.e., 
\begin{align}
	\bs^G&= \int_V \rho \bx \left[ \Delta\bOmega\times \bx \right] dV \quad\text{or} \nonumber\\
	&=\int_V \rho \bx \left[\bQ\cdot{\left(\bx_C \times \bJ \right)} \times \bx \right] dV,
\end{align}
where the superscript $G$ stands for the geometric part of $\bs$.
Hence
\begin{align}
\bs = \br_M \bJ + \bs^G.
\end{align}

In tensor notation, $\bs^G$ is:
\begin{align}
	s_{ij}^G &= N_{il} \epsilon_{jkl} \Delta\Omega_{k} \nonumber\\
	& = \epsilon_{jkl} N_{il} \left[\bQ\cdot{\left(\bx_C \times \bJ \right)}\right]_k.
\end{align}
Here $\epsilon_{jkl}$ is the Levi-Civita permutation symbol.

Up to this point, the derivation is for one rigid body in the collision pair. 
Due to symmetry and Newton's third law, the collision stress generated by this pair of particles, $1$ and $2$, is simply:
\begin{align}
\label{eq:colstressgeneral}
	\sigma_{ij}^{12} &= \left(r_{M,i}^2 -r_{M,i}^1 \right) F_j^{C} \nonumber \\
	&+ \epsilon_{jkl} N_{il}^2 \left[\bQ^2\cdot{\left(\bx_C^2 \times \bF_C \right)}\right]_k \nonumber \\
	&+ \epsilon_{jkl} N_{il}^1 \left[\bQ^1\cdot{\left(\bx_C^1 \times \bF_C \right)}\right]_k.
\end{align}
Here $\bF_C$ points from particle 1 to particle 2.

Again, the first term in Eq.~(\ref{eq:colstressgeneral}) is simply the virial stress, computed with the center of mass of the two particles.
The extra terms are contributions due to the particles' shape, mass distribution, etc.
For objects with homogeneous density $\rho$, the formula, Eq.~(\ref{eq:colstressgeneral}), is purely geometric, because the density $\rho$ in $N_{il}$ and $\bQ=\bG_M^{-1}$ cancel.
Also, since the equations of motion, Eqs.~(\ref{eq:rbeomtrans}) and~(\ref{eq:rbeomrot}), are linear, the stress generated by multiple collisions between two particles, or several particles colliding with one particle, can all be simply summed over each $\bF_C$.

In the above derivation, we made no assumption about how $\bF_C$ is computed. In general, $\bF_C$ can be computed in many different ways, depending on the physical setting and the collision resolution algorithms.
For example, for simple smooth spheres $\bF_C$ can be computed with WCA potentials.
While for more realistic granular flow models\cite{Campbellstresstensortwodimensional1986}, $\bF_C$ can be computed with considerations for having coefficient of restitution and friction.
The derivation of Eq.~(\ref{eq:colstressgeneral}) is straightforward but surprisingly not appreciated in the literature, except for a few special cases which we will show that Eq.~(\ref{eq:colstressgeneral}) reproduces those results.

\subsection{Mechanical pressure of $\sigma_{ij}^{12}$.}
The mechanical pressure is defined as the isotropic diagonal part of the stress. For $\sigma_{ij}^{12}$ given by Eq.~(\ref{eq:colstressgeneral}), we can show that:
\begin{align}
	\delta_{ij}\sigma_{ij}^{12} = \delta_{ij} \left(r_{M,i}^2 -r_{M,i}^1 \right) F_{C,j}.
\end{align}
In other words, the extra geometric part of $\sigma_{ij}^{12}$ changes only the deviatoric part of the collision stress. This is because $\delta_{ij}\epsilon_{jkl} N_{il} \Omega_k = \epsilon_{jkl} N_{jl} \Omega_k =0$, for any $ \Omega_k$, due to the symmetry of $N_{jl}$ and antisymmetry of $\epsilon_{jkl}$.

Therefore the mechanical collision pressure follows the usual virial formula:
\begin{align}
	\Pi^{12}=\frac{1}{3}\left(r_{M,i}^2 -r_{M,i}^1 \right) F_{C,i}.
\end{align}
% This is partially the reason why the pressure reported in different research are probably all correct although different equations may be used to compute $\sigma_{ij}^{12}$.

\subsection{Homogeneous frictionless spheres}
In the case of homogeneous frictionless spheres, we always have $\bF_C \parallel (\br_M^1 -\br_M^2) \parallel\bx_C^1 \parallel \bx_C^2$.
Also $\br_M$ coincides with the geometric sphere center due to homogeneity.
Therefore the geometric contribution to stress is zero, and we have the usual virial formula:
\begin{align}
	\sigma_{ij}^{12} &= \left(r_{M,i}^2 -r_{M,i}^1 \right) F_{C,j}^{1},
\end{align}
as has been widely used in many studies on the rheology of spherical suspensions \cite{foss_brownian_2000,wang_constant_2015}.

\subsection{Homogeneous frictional spheres}
In the case of homogeneous frictional spheres, the collision force $\bF_C$ is applied at the point of contact between the two spheres. In the special case of two equal spheres, we have $\bx_C^1=-\bx_C^2$, and Eq.~(\ref{eq:colstressgeneral}) reduces to:
\begin{align}
	\label{eq:frictionalsphere}
\sigma_{ij}^{12} &= \left(r_{M,i}^2 -r_{M,i}^1 \right) F_{C,j}^{1}.
\end{align}
However, unlike the frictionless case, $\bF^C$ is not necessarily parallel to $\br_M^2-\br_M^1$.
Equation~(\ref{eq:frictionalsphere}) reproduces the formula used by \citet{Campbellstresstensorsimple1989}.
% In general polydispersed cases, the extra geometric terms in Eq.~(\ref{eq:colstressgeneral}) do not necessarily vanish.

\subsection{Homogeneous frictionless long and thin rod}
In the case of homogeneous frictionless long and thin rod, the shape and orientation of each body is solely determined by an orientation norm vector $\bn$.
Taking the rod simply as a line segment, any point $\bx$ on the rod can be specified by:
\begin{align}
	\bx = x\bn ,\quad x\in[-L/2,L/2].
\end{align}
In this case, head-to-head collision is negligible because of the assumption of being long and thin.
Then in the absence of friction we always have $\bJ \perp \bn$. Therefore $\Delta\bOmega = ({\bx_C\times \bJ })/{\gamma}$, with $\gamma = \rho \int_{-L/2}^{L/2} x^2 dx$, and we have
\begin{align}
	\bs^G&= \frac{\rho}{\gamma} x_C \bn\bJ\int_{-L/2}^{L/2} x^2 dx =\bx_C \bJ.
\end{align}
Further, Eq.~(\ref{eq:colstressgeneral}) reduces to:
\begin{align}
	\sigma_{ij}^{12} &= \left(r_{M,i}^2 + x_{C,i}^2 -r_{M,i}^1 - x_{C,i}^1 \right) F_{C,j}^{1},
\end{align}
which reproduces the formula used in the work by \citet{SnookNormalstressdifferences2014}.

\section{Collision resolution in dynamic simulations}
\label{sec:colalgo}
The other ingredient in our calculation of the collision stress is how to stably and efficiently compute the collision force $\bF_C$ needed for Eq.~(\ref{eq:colstressgeneral}).
For underdamped systems with inertia, significant progress have been made by \citet{tasora_large-scale_2008}.
In this work we extend this approach to overdamped systems, because most active matter systems we are interested in are in this regime.
Accordingly, we also focus on the completely inelastic collision case, where colliding bodies can remain in contact after collisions.
Here we ignore friction.

\subsection{The mobility problem}
We start from the mobility problem because having the mobility matrix being symmetric-positive-definite (SPD) is one of the keys to the success of our method.
Due to the linearity of Stokes equation, the dynamics of $n_b$ rigid bodies is specified compactly by a linear equation:
\begin{align}
\bUcal = \bMcal \bFcal, \label{eq:mobmat}
\end{align}
where $\bUcal=(\bU_1,\bOmega_1,\bU_2,\bOmega_2,...)$ consists of translational and rotational velocities of each rigid body,
and $\bFcal=(\bF_1,\bT_1,\bF_2,\bT_2,...)$ consists of the forces and torques on each rigid body. They are both column vectors with $6n_b$ entries.
$\bMcal$ is the \emph{mobility matrix}, which contains all the solution information given by the Stokes equation and the no-slip boundary condition.
% Equivalently, we can define a \emph{resistance matrix} $\bRcal$ such that $\bFcal = \bRcal \bUcal$, and $\bRcal = \bMcal^{-1}$.
That the mobility matrix $\bMcal$, and consequently the resistance matrix $\bRcal=\bMcal^{-1}$, is SPD is well-known \cite{Kim_Karrila_2005}.
Physically, the positive-definiteness can be explained by a simple observation, that any non-zero force $\bFcal$ applied to the rigid bodies dissipates energy into the viscous fluid, that is,
\begin{align}
	\bFcal\cdot\bUcal = \bFcal^T \bMcal\bFcal >0.		
\end{align}

It is important that all the derivations in this work make no assumption about the shape of the rigid bodies, nor of the numerical method used to solve the mobility problem.
Also, our approach does not require that the matrix $\bMcal$ be explicitly constructed.
As long as $\bUcal$ can be computed with given force $\bFcal$ for a given geometry, the method derived in this work can be applied.
At the most crude level of description, the many-body coupling can be completely ignored and $\bMcal$ becomes block-diagonal, describing isolated Brownian particles.
With many-body coupling, the Rotne-Prager-Yamakawa tensor is a fairly inexpensive SPD approximation to $\bMcal$, and can be used here straightforwardly.
Stokesian Dynamics\cite{wang_spectral_2016} can also be used in this method as a full hydrodynamics solver.
The recent progress in boundary integral methods provides the most accurate solvers to the mobility problem, for which spheres \cite{corona_integral_2017,corona_boundary_2018} and rigid slender bodies \cite{tornberg_numerical_2006,gustavsson_gravity_2009} are examples.

\subsection{Complementarity formulation for contact dynamics}
The evolution of the geometric configuration $\bq$ of a collection of rigid bodies is uniquely defined by the translational and rotational velocities $\bU_k$ and $\bOmega_k$ for each particle $k$. 
Their velocities can be partitioned as the `known' velocities, and the `collision' velocities:
\begin{align}
\bU_k&=\bU_{k,known}+\bU_{k,C}, \\
\bOmega_k&=\bOmega_{k,known}+\bOmega_{k,C},
\end{align}
where `known' stands for the known velocities before resolving the collisions.
For example, for Brownian colloids, $\bU_{k,known}$ and $\bOmega_{k,known}$ are Brownian displacements which can be computed without resolving the consequent collisions.
Also for swimming bacterial $\bU_{k,known}$ and $\bOmega_{k,known}$ arise from the swimming motion.

The collision motion $\bUcal_C=\bMcal \bFcal_C$ is governed by the mobility problem Eq.~(\ref{eq:mobmat}).
The collision velocities $\bUcal_C$ are governed by the mobility problem $\bUcal_C = \bMcal_C\bFcal_C$, i.e., Eq.~(\ref{eq:mobmat}).
The equations of motion for the rigid bodies can be written as the evolution of configuration $\bq$ with velocity $\bUcal$:
\begin{align}
\dot{\bq} &=  \bUcal, \label{eq:particle_evolution} \\
\bUcal &= \bUcal_{known} + \bMcal \bFcal_{C}. \label{eq:force_balance}
\end{align}
In this formulation, both $\bFcal_C$ and $\bUcal_C$ are the unknowns to be solved for, with the geometric constraint that $\bq$ satisfies the non-overlap condition at the end of each timestep.
The geometric non-overlap condition can be defined as having a positive minimal separation, that is, $\Phi_\ell(\bq)>0$ between each close pair $\ell$ of rigid bodies, as a function of geometry configuration $\bq$.

For each contact pair indexed $\ell$, the positivity of minimal separation distance $\Phi_\ell$ and the collision force magnitude $\gamma_\ell$ are mutually exclusive situations:
\begin{itemize}
\item No contact: $\Phi_\ell>0$ and $\gamma_\ell=0$.
\item Contact: $\Phi_\ell=0$ and $\gamma_\ell>0$.
\end{itemize}
Mathematically this is called a complementarity condition, and is usually denoted by the following special notation combining all $\ell$:
\begin{align}\label{eq:cpproblem}
	\CP{\bPhi}{\bgamma},
\end{align}
where $\bPhi=(\Phi_0,\Phi_1,...)$ denotes the collection of minimal distances, and $\bgamma=(\gamma_0,\gamma_1,...)$ denotes the collection of all contact force magnitudes, for all possible contacts in the system.
The dimension of both $\bPhi$ and $\bgamma$ is $n_C$, the total number of possible collisions in the system.
$n_C$ is identified by tracking the separation distance between pairs of rigid bodies that are close to collision. 
That is, once a pair of particles' separation $\Phi_\ell$ is larger than a positive distance $\delta$, this pair is then excluded from the collision resolution algorithm because they are far apart and cannot collide within one timestep.
This threshold distance $\delta$ is chosen empirically according to the system dynamics, and is not necessarily a constant for all pairs or all timesteps.
For example, we usually pick $\delta = 0.5(R_i+R_j)$  for a pair of spheres with radius $R_i$ and $R_j$. 

Now, for $n_b$ rigid bodies appearing in the mobility problem, let $\bD_\ell \in \mathbb{R}^{6 n_b}$ be a sparse column vector containing geometric information mapping the magnitudes $\gamma_\ell$ to the collision force (and torque) vector on each rigid body.
$\bD_\ell$ defined in this way gives the force and torque on the two rigid bodies in this collision pair $\ell$, as a linear function to the collision force magnitude $\gamma_\ell$.
Therefore $\bD_\ell$ has 12 non-zero entries for aspherical shapes, corresponding to 3 translational and 3 rotational degrees of freedom for each rigid body in the contact pair.
For two spheres in contact without friction, $\bD_\ell$ has only 6 non-zero entries because the normal collision forces induces no torques in this case.
Then we can define a matrix $\bDcal\in\bbR^{6 n_b \times n_C}$ as the assembly of all $\bD_\ell$ column vectors, mapping $\bgamma$ to the collision forces $\bFcal_C$:
\begin{align}
	\bFcal_C &=\bDcal\bgamma,\\
	\bDcal &=\left[\bD_0\,\bD_1\,\ldots\,\bD_{n_C} \right] \in \mathbb{R}^{6 n_b \times n_C}.
\end{align}
The details about entries of $\bDcal$ can be found in the work by \citet{tasora_large-scale_2008}.
% For general asphericalal rigid bodies colliding with a boundary, $\bD_\ell$ also has only 6 non-zero entries because in this work a boundary is defined as an object fixed or moving with known velocity, and therefore does not appear in the mobility matrix $\bMcal$ to be applied to $\bFcal_c$.

Then, the equations of motion result in the  differential variational inequality
\begin{align}
\dot{\bq} &= \bUcal(\bq),\\
\bUcal(\bq) &= \bUcal_{known}(\bq) +  \bMcal(\bq)  \bDcal(\bq) \bgamma, \\
&\CP{\bPhi(\bq)}{\bgamma}.
\end{align}
Here $\bUcal_{known}(\bq)$, $\bMcal(\bq)$, and $\bDcal(\bq)$ are all directly solvable with given geometry $\bq$, without information about the collision force magnitudes $\bgamma$.
This equation set is then solvable and integrable in time once a relation between the configuration $\bq$ and the collision force $\bgamma$ is supplied, that is, a timestepping scheme.
Higher order schemes such as the Runge-Kutta and Adams-Bashforth families can all be used, but for simplicity of derivation we employ a first-order Euler scheme.
Given position $\bq^k$ and velocity $\bUcal_{known}^k$ at a given time step $t^k$ and step size $\Delta t$, velocity $\bUcal^{k}$ and contact forces $\bgamma^{k}$ are solved via the nonlinear complementarity problem (NCP):
\begin{align}
\bq^{k+1} &= \bq^k + \Delta t \left(\bUcal_{known}^k +  \bMcal(\bq)^k  \bDcal^k \bgamma^k\right), \\
&\CP{\bPhi(\bq^{k+1})}{\bgamma^k}.
\label{eq:stepncp}
\end{align}
The velocity $\bUcal^{k}$ is then used to evolve the position in time.

This is an NCP because the minimum gap $\bPhi$ is in general a nonlinear function of $\bq$.
NCPs can often be solved iteratively by a series of linear complementarity problems (LCP) with superlinear or quadratic convergence rate \cite{fang_linearization_1984}.
Here we follow a simpler route rather than solving the NCP exactly. 
The timestep size $\delta t$ must be reasonably small to integrate $\bUcal_{known}^k$ accurately, and so $\bPhi(\bq^{k+1}) \geq 0$ can be linearized (and scaled with $1/\Delta t$) to yield:
\begin{align}
\frac{1}{\Delta t}\bPhi(\bq^{k}) + \left(\nabla_{\bq}\bPhi\right)^k\left[ \bUcal_{known}^k + \bMcal^k \bDcal^k\bgamma^k \right]\geq 0,
\end{align}
where the matrix $\left(\nabla_{\bq}\bPhi\right)^k$ is simply the coefficients of the Taylor expansion of $\bPhi$ over $\bq$ at timestep $t^k$.

For rigid objects, it is straightforward to show that $\nabla_{\bq}\bPhi = \bDcal^T$.
This is the same relation utilized in the work by \citet{tasora_large-scale_2008}.
The LCP problem can be written in the standard form:
\begin{align}
\CP{\bA^k\bgamma^k+\bb^k}{\bgamma^k},\label{eq:LCPdef}
\end{align}
where
\begin{align}
\bA &= \bDcal^T \bMcal\bDcal, \label{eq:LCPdefA} \\
\bb &=\frac{1}{\Delta t}\bPhi(\bq) + \bDcal^{T} \bUcal_{known}.\label{eq:LCPdefb}
\end{align}
The term $\bDcal^{T} \bUcal_{known}$ computes the (linearized) changes in the minimal separation $\bPhi$ before the contact constraints are considered.
We also note that each application of $\bA$ corresponds to the solution of a mobility problem for the contact force $\bFcal_C = \bDcal\bgamma$.
For large enough numbers of particles, it may thus be preferable to use matrix-free methods instead of constructing $\bA$ explicitly.

The procedures of this collision resolution method based on LCP are:
\begin{itemize}
	\item[1.] Compute $\bUcal_{known}^k$ at timestep $t^k$.
	\item[2.] Compute the sparse matrix $\bDcal^k$ with given geometric configuration $\bq^k$ and the threshold $\delta$ for possible contacts.
	\item[3.] Solve for $\bgamma^k$ with Eq.~(\ref{eq:LCPdef}). $\bUcal_C^k$ and $\bFcal_C^k$ are solved simultaneously.
	\item[4.] Evolve to $\bq^{k+1}$ with $\bUcal_{known}^k+\bUcal_C^k$.
\end{itemize}

\subsection{LCP solvers}
In this section we briefly discuss the solution methods to Eq.~(\ref{eq:LCPdef}).
The superscripts $k$ denoting the timestep are dropped to simplify the notation, since the LCP solution algorithms discussed here are generic methods not limited to collision resolution problems.

% \subsubsection*{First order methods}
The matrix $\bA$ defined in the LCP formulation Eq.~(\ref{eq:LCPdef}) is symmetric-positive-semi-definite (SPSD), because the mobility matrix $\bMcal$ is symmetric-positive-definite (SPD).
Therefore the LCP problem can be conveniently converted to a Constrained Quadratic Programming (CQP) \cite{niebe_numerical_2015}:
\begin{align}\label{eq:CQPdef}
 \bgamma = \arg\min_{\bgamma\geq 0} f(\bgamma) = \frac{1}{2} \bgamma^T \bA \bgamma + \bb^T \bgamma.
\end{align}
From the physics perspective, the minimization of $f(\bgamma)$ can be understood qualitatively as the minimization of the total virtual work done by the collision forces (and torques).
This CQP formulation allows a wide range of algorithms.
It can be solved with first order methods based on Projected Gradient Descent (PGD), where the projection is used to impose the constraint $\bgamma\geq0$ during the gradient-descent minimization process.
It can also be solved with second order Newton-type methods, for example, the minimum-map Newton method\cite{niebe_numerical_2015}.

It is beyond the scope of this work to discuss these methods in detail.
Here we solve the LCP problem with first order methods, because we found PGD methods are much more efficient since the gradient $\bg=\nabla f = \bA\bgamma+\bb$ is inexpensive to compute for every gradient descent step.
In particular, we found that Barzilai-Borwein Projected Gradient Descent (BBPGD) is much more efficient than the previously reported Accelerated Projected Gradient Descent (APGD) \cite{mazhar_using_2015},
because BBPGD does not rely on the estimation by back-tracking of the Lipschitz parameter of the function $f$.
The BBPGD algorithm has been analyzed mathematically for generic CQP by \citet{dai_projected_2005}.
The procedures of BBPGD can be found in Appendix~\ref{sec:appbbpgd}.

The convergence of CQP solvers can be checked at each step by computing the $L_2$-norm $\phi(\bgamma,\bg(\bgamma))$ of the minimum-map function $\bH$:
\begin{align}
\phi(\bgamma,\bg(\bgamma)) &= \NORMtwo{\bH(\bgamma,\bg(\bgamma))} < \epsilon_{tol},\\
\bH(\bgamma,\bg(\bgamma)) &= \min(\bgamma,\bA\bgamma+\bb),
\end{align}
because the solution to the CQP is reached when $\phi=0$.
In this work, $\epsilon_{tol}=10^{-5}$ is used unless otherwise noted.
This criteria function $\phi$ is also efficient to compute because $\bA\bgamma+\bb$ is already computed as the gradient of the quadratic function at each gradient descent step.

\subsection{Performance}
The collision resolution algorithm based on the LCP Eq.~(\ref{eq:LCPdef}), allows the timestep size $\Delta t$ to be increased by 10 $\sim$ 100 times in comparison to the traditional method with LJ or WCA potentials, because the stiffness induced by the potentials is eliminated. 
For each timestep, the explicit construction of Eq.~(\ref{eq:LCPdef}) has approximately the same cost as computing the pairwise repulsive force.
After the construction, $\bA\bgamma+\bb$ must be computed once during each BBPGD minimization step. 
The total number of iterations increases slowly with the number of actual collisions, i.e., the number of positive entries in the solution $\bgamma$.
Empirically, $5\sim10$ iterations is enough for dilute systems.
Since $\bA\bgamma+\bb$ can be computed with standard sparse matrix-vector multiplication operations (\texttt{spmv}) efficiently, the solution of Eq.~(\ref{eq:LCPdef}) is usually not a significant extra cost unless the system is densely packed and close to the random-close-packing (RCP) limit, where $O(1000)$ iterations is necessary.
Therefore overall this LCP-based method significantly increases both the stability and efficiency of resolving collisions compared to repulsive potential methods. 
We also implemented this algorithm with full \texttt{MPI} and \texttt{OpenMP} parallelism, and the program scales efficiently to $O(10^7)$ particles on $O(100)$ cores.

\section{Validation}
\label{sec:benchsylinder}
To validate our derivation for the collision stress and resolution algorithm, particles with aspherical shapes should be used because otherwise the geometric part in Eq.~(\ref{eq:colstressgeneral}) varnishes. 
Unfortunately such widely accepted and available benchmark data is only available for a few systems, partially because the difficulty of handling collisions between aspherical particles and computing the stress.

In this section we extract the Equation-of-State (EOS) of monodisperse Brownian spherocylinders of length $L$ and diameter $D$, 
and compare the results with the benchmark data reported by \citet{bolhuis_tracing_1997}.
In this purely Brownian system, the many-body hydrodynamics coupling in the mobility matrix is ignored, i.e.,
$\bMcal$ becomes block diagonal, with each block being the translational and rotational mobility matrix $\bM_{tt}$ and $\bM_{rr}$ for each spherocylinder.
The coupling between rotational and translational motion is also ignored:
\begin{align}
	\bM_{tt} & = \frac{1}{\zeta_\parallel} \bn\bn^T + \frac{1}{\zeta_\perp}\left(\bI-\bn\bn^T\right),\\
	\bM_{rr} &= \frac{1}{\zeta_r}\bI.	
\end{align}
Here $\bn$ is the orientation norm vector of the spherocylinder.
The drag coefficients are approximated by slender body theory of straight rigid fibers \cite{tornberg_numerical_2006}:
\begin{align}
1/\zeta_\parallel &= 2 b / (8 \pi L \mu ) ,\\	
1/\zeta_\perp &= (b+2) / (8 \pi L \mu ),\\	
1/\zeta_r &= 3(b+2) / (2 \pi L^3 \mu ),\\	
b &= -\left(1 + 2 \log[{D}/({2L})]\right).
\end{align}
Similar but different drag coefficients are often used in previous work \cite{lowen_brownian_1994,tao_brownian_2005-1}.
% However, the measured pressure is not impacted by the choice of the drag coefficients, as long as the relaxation time associated with $\zeta_\parallel$, $\zeta_\perp$, $\zeta_r$ are reasonable, because these drag coefficients are only used to generate the translational and rotational Brownian motion, and will be scaled out by $k_BT$.

\begin{figure}[htb]
\includegraphics[width=\linewidth]{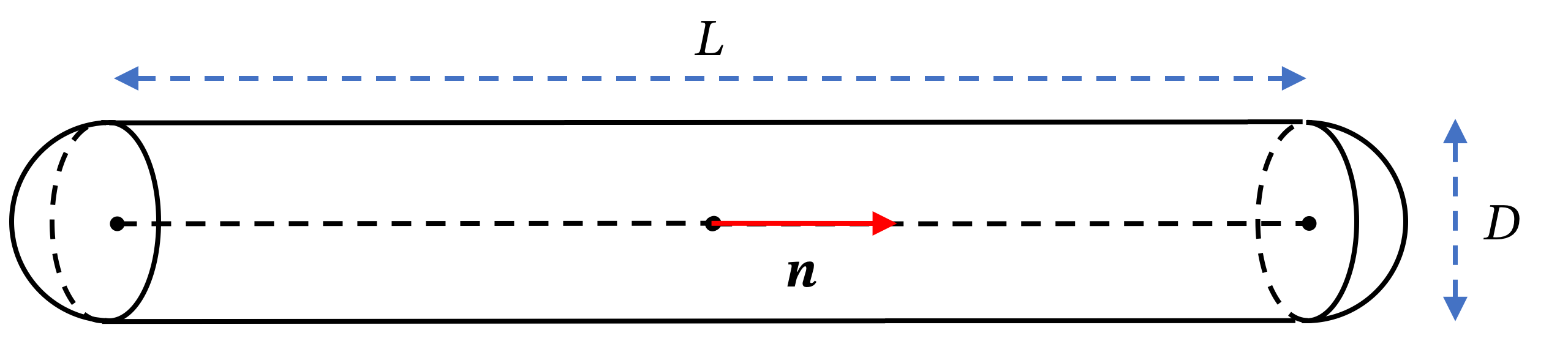}	
\caption{\label{fig:geosylinder} The geometry of a spherocylinder of length $L$, width $D$, and orientation $\bn$.}
\end{figure}

The Brownian velocities $\bU^B$ and $\bOmega^B$ for each spherocylinder are computed by the Random-Finite-Difference (RFD) algorithm\cite{delong_brownian_2015} treating $\bM_{tt}$ and $\bM_{rr}$ independently, because the many-body coupling has been ignored.
Then the `known' velocity $\bUcal_{known}$ in Eq.~(\ref{eq:force_balance}) is just the Brownian velocity  $\bUcal^B=(\bU_1^B,\bOmega_1^B,\bU_2^B,\bOmega_2^B,...)$.
The necessary geometric quantities in Eq.~(\ref{eq:colstressgeneral}) and the sparse matrix $\bDcal$ in the LCP collision resolution algorithm are computed with the method described in Appendix~\ref{sec:appsylinder} and~\ref{sec:appcolsylinder}. 
BBPGD algorithm is then used to solve the CQP (equivalent to the LCP) for $\bUcal_C$ and $\bFcal_C$.
The system stress is then computed with $\bFcal_C$ according to Eq.~(\ref{eq:colstressgeneral}) for each pair in the collision.

Periodic boundary conditions are imposed in each direction of the rectangular simulation box of size $L_x\times L_y\times L_z$, containing $N$ spherocylinders. $n=N/(L_xL_yL_z)$ is the number density.
The system total stress and pressure are computed with a simple average of all collision pairs' contributions:
\begin{align}
	\bSigma &= n k_BT \bI + \frac{1}{N}\sum \bsigma^{col}	,\\
	\Pi &= \frac{1}{3}\mathrm{Tr}\bSigma,\quad \Pi^{col} = \frac{1}{3N}\sum\mathrm{Tr} \bsigma^{col}.
\end{align}
The kinetic part $n k_B T \bI$ is \emph{imposed} with given $k_BT$ through the Brownian motion moves in overdamped simulations.

\subsection{The isotropic phase}
\begin{figure}[htbp]
	\includegraphics[width=\linewidth]{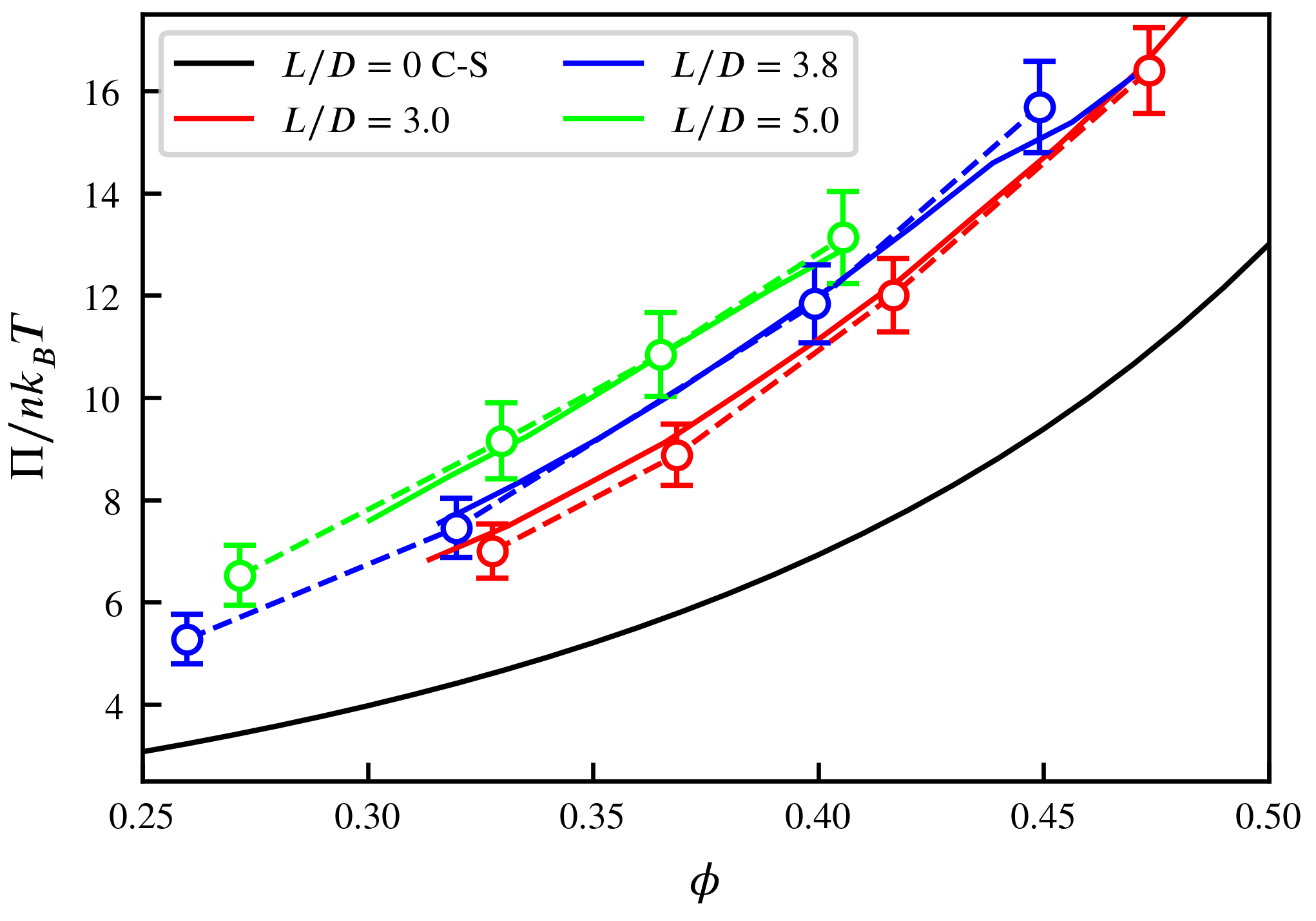}	
	\caption{The pressure of Brownian spherocylinders in the isotropic phase. The open circles with error bars connected by dashed lines are measured from simulations. 
	The error bars show the standard deviation of pressure within the time-average window. 
	The solid lines are data extracted from the work by \citet{bolhuis_tracing_1997}. 
	The black line shows the Carnahan-Starling equation for hard spheres ($L/D=0$) as a reference. \label{fig:SylinderEos}}
\end{figure}

We first present results in the isotropic phase, as shown in Fig~\ref{fig:SylinderEos}.
The simulations start from a random placement and orientation of $N=2000$ spherocylinders of varying aspect ratio $L/D$ in a cubic periodic box, 
and are equilibrated with fixed box size until the measured stress reaches a steady state.
This process usually takes about $10^5$ timesteps. 
Then the system pressure $\Pi$ is averaged over another $2000$ timesteps. 
The method described in this work accurately reproduces the standard data reported by~\citet{bolhuis_tracing_1997}.

\subsection{The isotropic-nematic phase transition}
Beyond the isotropic phase, the simulations are much more demanding because the system relaxation time becomes significantly longer. 
In this regime, if a simulation is simply started from a random configuration, it remains `jammed' in this structure for a long time, even when the system density is in the nematic phase regime.
Limited by computing resources, we conduct dense simulations starting from $N$ randomly located, but all aligned configuration of spherocylinders.
The fixed simulation box is fixed with $L_x>L_y=L_z$, and the spherocylinders are aligned in the $x$ direction. 
$N=2000$ is fixed but the box sizes are varied around $72D\times15D\times15D$ for different volume fractions. 
Simulations with $N=6000$ spherocylinders in a cubic periodic box are also performed and the results reported here are not impacted by the box shape.

\begin{figure}[htbp]
	\includegraphics[width=\linewidth]{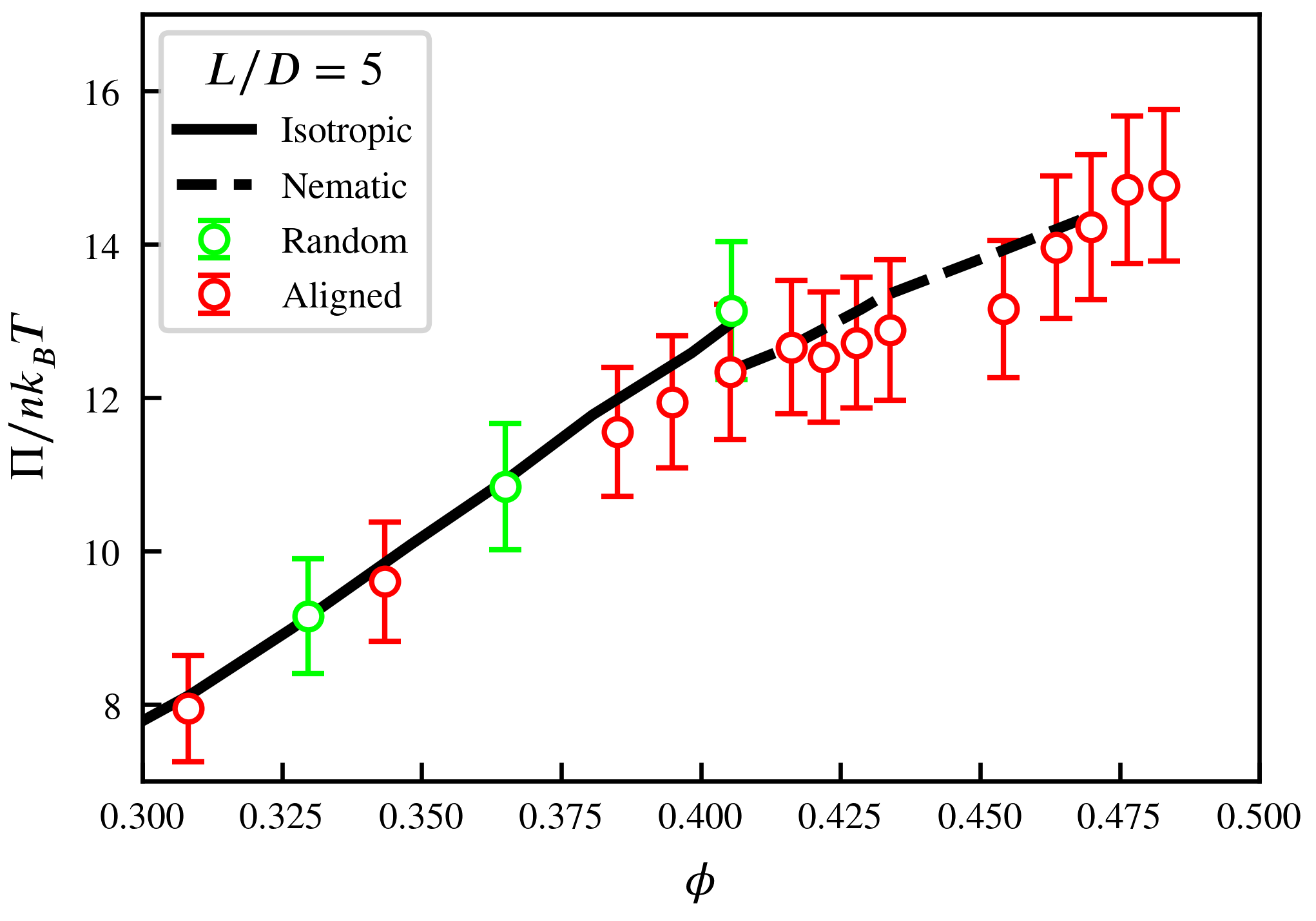}	
	\caption{
		The pressure of Brownian spherocylinders close to the isotropic-nematic phase transition. 
		The green symbols are simulations starting from a randomly oriented configuration in a cubic box, and the red symbols are simulations starting from a random center location but aligned orientation in a rectangular box. 
		The error bars show the range of standard deviation of pressure within the time-average window. 
		The solid line representing the isotropic phase, and the dashed line, representing the nematic phase, are both data extracted from the work by \citet{bolhuis_tracing_1997}. \label{fig:IsoNemLD50}}
\end{figure}

We focus on the isotropic-nematic transition for $L/D=5$, where a nematic phase can stably exist, because it is not too close to the isotropic-nematic-smectic triple point at around $L/D\approx 3.7$ estimated by \citet{bolhuis_tracing_1997}. 
The pressure and its standard deviation is also calculated with equilibrated systems in the same way as described above. 
The results for the measured pressure agrees well with the results by \citet{bolhuis_tracing_1997}, as shown in Fig.~\ref{fig:IsoNemLD50}.

\begin{figure}[htbp]
	\includegraphics[width=\linewidth]{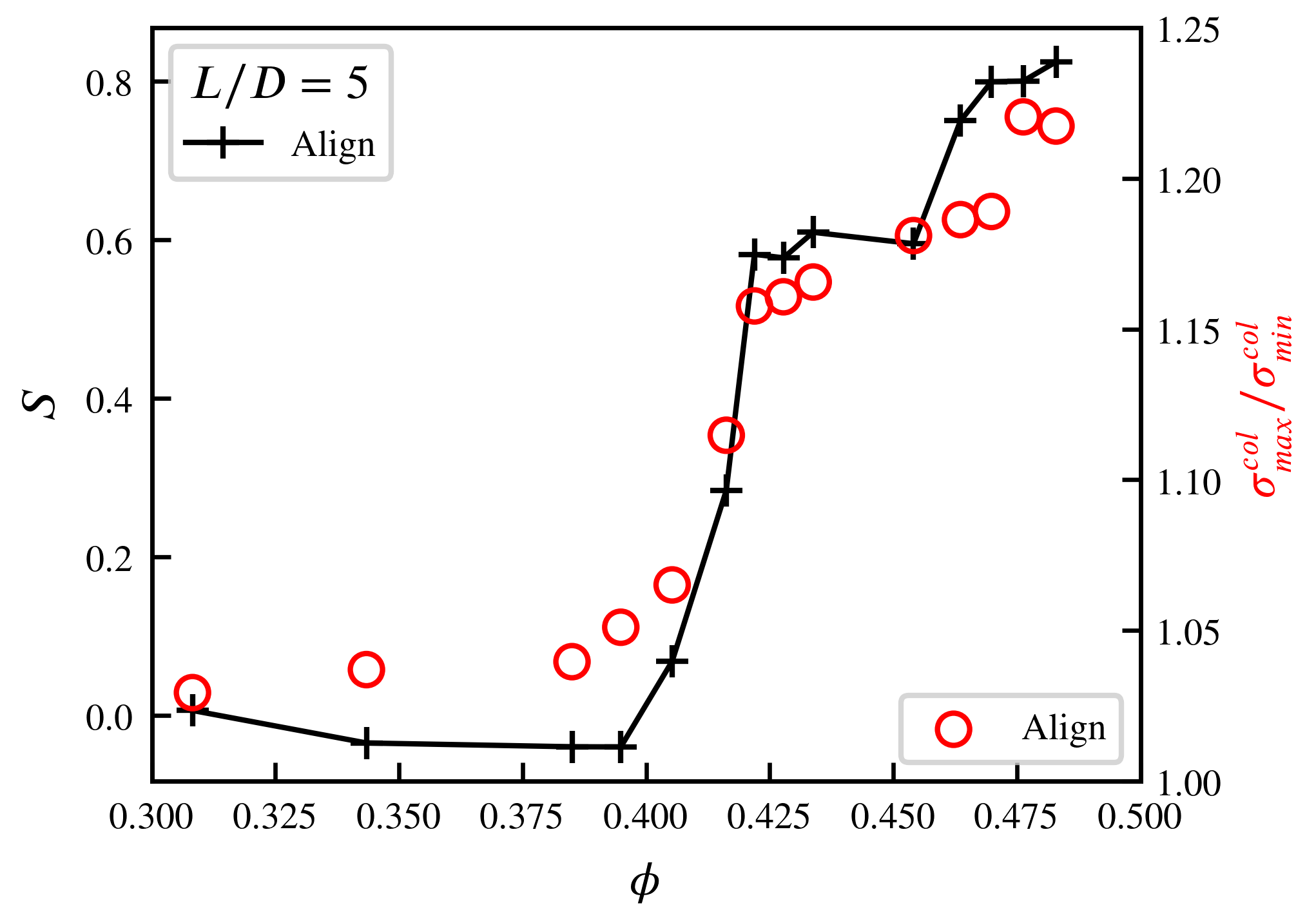}	
	\caption{The jump in orientation order parameter $S$ and the collision stress anisotropy during the isotropic-nematic phase transition for $L/D=5$. 
	The black symbols connected by a solid line shows the order parameter $S$, and the red open circles show the anisotropy. 
	All data are extracted from the same set of simulations starting from a random but aligned state, as in Fig.~\ref{fig:IsoNemLD50}. \label{fig:SorderLD50}}
\end{figure}

Fig.~\ref{fig:IsoNemLD50} shows a jump in pressure at $\phi\approx0.4$.
More information about this isotropic-nematic transition can be extracted by measuring the orientation order parameter $S=\AVE{P_2(\bn\cdot\bar{\bn})}$, where $P_2$ is the order-$2$ Legendre polynomial, and $\bar{\bn}$ is the average orientation of spherocylinders at a specific time in the simulation.
Further, the anisotropy of the system pressure can be quantitatively investigated by computing the ratio of the maximum to the minimum of the eigenvalues $\sigma_{max}^{col}/\sigma_{min}^{col}$ of the collision stress tensor $\bsigma^{col}$. 
As shown in Fig.~\ref{fig:SorderLD50}, the anisotropy ratio $\sigma_{max}^{col}/\sigma_{min}^{col}$ closely follows the jump in $S$, which shows the isotropic-nematic phase transition for $L/D=5$ happens at $\phi\approx0.42$.
Last, but not least, the computed stress tensor $\bsigma^{12}$ is exactly symmetric without Brownian noise for each pair of spherocylinders at each timestep, as required by the general principal  of continuum mechanics.
This would not be satisfied if the geometric part in Eq.~(\ref{eq:colstressgeneral}) is not included in the stress calculation.

\section{Application}
\label{sec:application}
In this section, we demonstrate a few applications of the computational framework described in this work to the area of soft active matter, namely, self-propelled rods and growing-dividing cells.

\subsection{Self-propelled rods}
\label{subsec:swimrod}
The Active Brownian Particle (ABP) model has attracted much attention because despite being a minimal model it can be used to explain many important features of soft active matter systems. 
However, the similar Self-Propelled Rod (SPR) model has not been investigated in such detail in the literature.
Almost all related work focuses on 2D systems \cite{baskaran_enhanced_2008,ginelli_large-scale_2010,orozco-fuentes_order_2013,kuan_hysteresis_2015-1,weitz_self-propelled_2015-1,peruani_active_2016,grosmann_mesoscale_2016}, mostly because the collisions are difficult to handle in 3D.
In particular, an EOS has not been quantitatively measured. 
In this work we report briefly on the enhancement of collision pressure for dilute  Brownian SPR systems. 
The Brownian SPR model we consider here is exactly the same as the Brownian spherocylinders considered in the last section, except that each spherocylinder has a propulsion speed $\bU_0$ along its orientation norm vector $\bn$. 

The virial expansion of the EOS can be written as: \cite{vroege_phase_1992}
\begin{align}
	\frac{\Pi}{nk_BT} = 1 + B_2 n + B_3 n^2 + \cdots,
\end{align}
	or,
\begin{align}
	1+\frac{\Pi^{col}}{nk_BT}  = 1+ B_2\frac{\phi}{v_0} + \frac{B_3}{B_2^2}\left(B_2\frac{\phi}{v_0}\right)^2 +\cdots,
\end{align} 
where $v_0=\pi\left(\tfrac{1}{4}LD^2+\tfrac{1}{6}D^3\right)$ is the volume of a single rod (spherocylinder).
In the limit of $\phi \to 0$, the higher order terms varnish and the EOS can be approximately written as:
\begin{align}
	\frac{\Pi^{col}}{nk_BT} \approx  B_2\frac{\phi}{v_0}.
\end{align}
When $U_0=0$, $B_2=\pi\left(\tfrac{2}{3}D^3+LD^2+L^2D/4\right)$ is analytically known \cite{onsager_effects_1949,graf_density_1999}. 
Therefore we measure the enhancement of collision pressure $\Pi^{col}$ due to self propulsion with simulations at a given $L/D$ and $\phi$, with varying $U_0$. 
We simulate $N=4\times 10^5$ SPRs in a fixed cubic periodic box to overcome the strong effect of Brownian noise in such dilute systems, and guarantee that the persistence length $U_0/D_R$ is much smaller than the box size.
We take $\phi=0.0052$ for $L/D=5$, $\phi=0.0065$ for $L/D=10$, and $\phi=0.0065$ for $L/D=20$.
Such dilute systems remain isotropic with varying $U_0$.
We plot the measured $\Pi^{col}/(B_2 n)$ as a function of dimensionless velocity $U_0/(LD_R)$, where $D_R=k_BT/\zeta_r$ is computed as in Section~\ref{sec:benchsylinder} when $\phi\to 0$. 

\begin{figure}[htbp]
	\includegraphics[width=\linewidth]{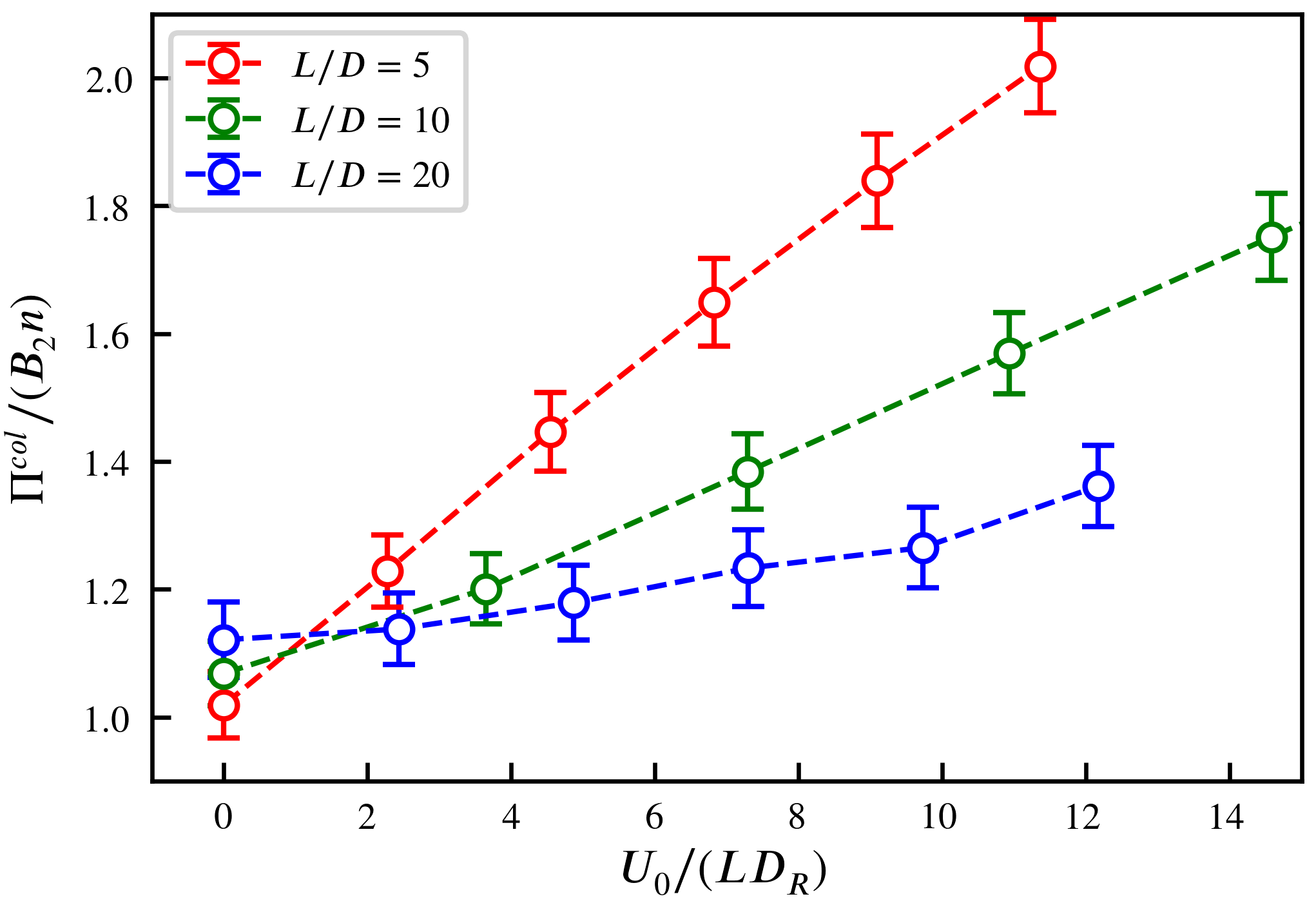}	
	\caption{The enhancement of collision pressure due to self-propelled velocity $U_0$ in the dilute limit. 
	Here $\phi=0.0052$ for $L/D=5$, $\phi=0.0065$ for $L/D=10$, and $\phi=0.0065$ for $L/D=20$. 
	The results and error bars are averaged for $2000$ timesteps over equilibrated systems. \label{fig:RodU0B2}}
\end{figure}

The results of this measurement is shown in Fig.~\ref{fig:RodU0B2}.
The collision pressure increases almost linearly as the propulsion speed $U_0$.
Some recent work\cite{kraikivski_enhanced_2006} proposed an `effective length' $L_U = \sqrt{L(L+U_0/D_R)}$  to approximate the effect of propulsion.
Substituting $L_U$ into the analytic expression for $B_2 =\pi\left(\tfrac{2}{3}D^3+LD^2+L^2D/4\right)$ does generate a linear scaling as $U_0$ when $L/D\to\infty$, but we found that quantitatively this simple scaling law fails in predicting both the value and the trends of the data shown in Fig.~\ref{fig:RodU0B2}.

Ideally, $\Pi^{col}/(B_2 n) \to 1$ at $U_0=0$, which is approximately the case of $L/D=5$. 
For $L/D=10$ and $20$, there is about $10\%$ error, because the contributions from $B_3,B_4$, etc., remain important. 
Using a more dilute system could help resolve this issue, but a larger number of SPRs are necessary to overcome the Brownian noise, which is currently beyond our computing power.
However, this slight mismatch does not change our conclusion of the linear scaling between $\Pi^{col}$ and $U_0$.

\subsection{Growing and dividing cells}
\label{subsce:gdcell}
\begin{figure}[htbp]
	\includegraphics[width=\linewidth]{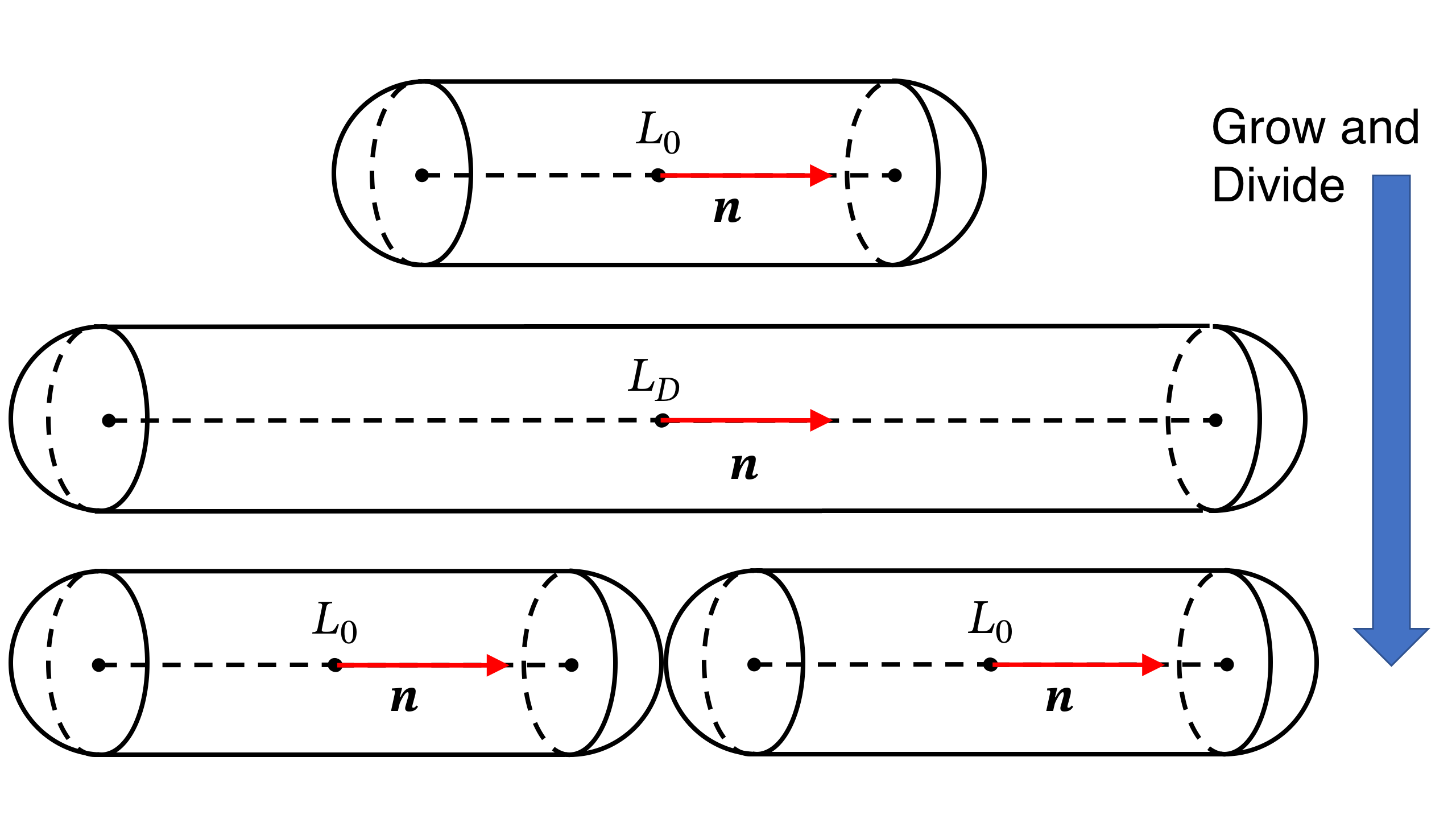}	
	\caption{The division of cells modelled as splitting of spherocylinders. The cell with length $L_D$ divides into two cells with equal length $L_0$. The total cell volume slightly decreases in this division process due to the shape change in the center. The orientation norm vector $\bn$ remains unchanged after the division.\label{fig:Division}}
\end{figure}
The collision stress Eq.~(\ref{eq:colstressgeneral}) and the LCP method Eq.~(\ref{eq:LCPdef}) are derived for rigid bodies in Section~\ref{sec:paircolstress} and~\ref{sec:colalgo}.
However, this assumption only means that they are rigid in response to collision forces.
Besides this, they can freely deform and both Eq.~(\ref{eq:colstressgeneral}) and Eq.~(\ref{eq:LCPdef}) are still applicable.
Growing and dividing cells are one of the examples with which we can demonstrate the applications where the objects are changing their shapes, even discontinuously.
In the following we present some interesting stress measurement for systems of a minimal model of growing and dividing cells. 
The model is unrealistic because the growing and diving process is assumed to be synchronized for all cells and the time between division is very short.
We use this model only to demonstrate the capability of the computational method. 
More realistic biological parameters can be straightforwardly added to this minimal model in our future study.

\begin{figure}[htb]
	\includegraphics[width=\linewidth]{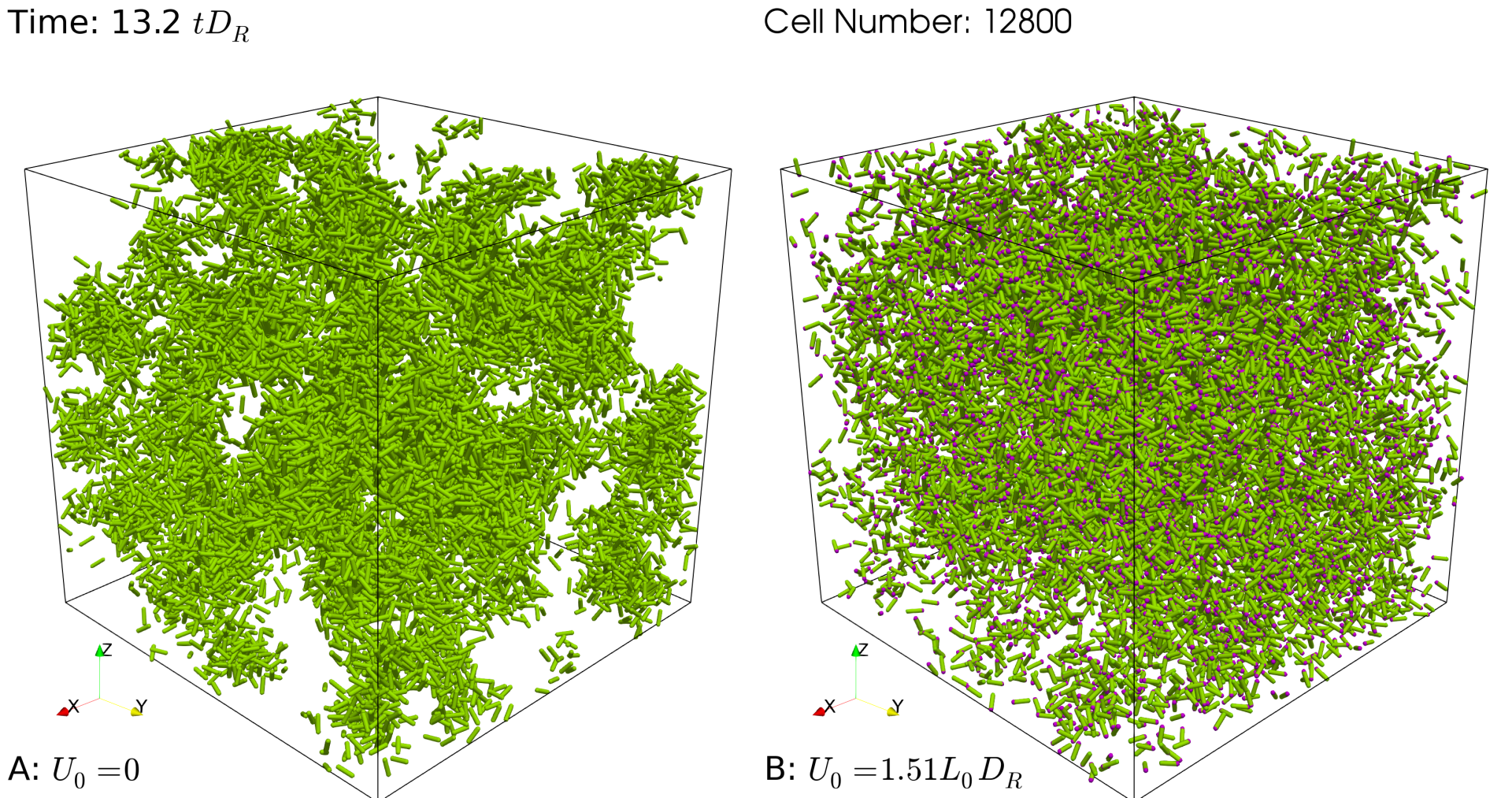}	
	\caption{The snapshot of dividing cells for A: $U_0=0$ and B: $U_0=1.51L_0D_R$ at time $tD_R = 13.2$. The purple dots mark the heads of the moving cells. \label{fig:SnapCellDivision250}}
\end{figure}

\begin{figure}[htb]
	\includegraphics[width=\linewidth]{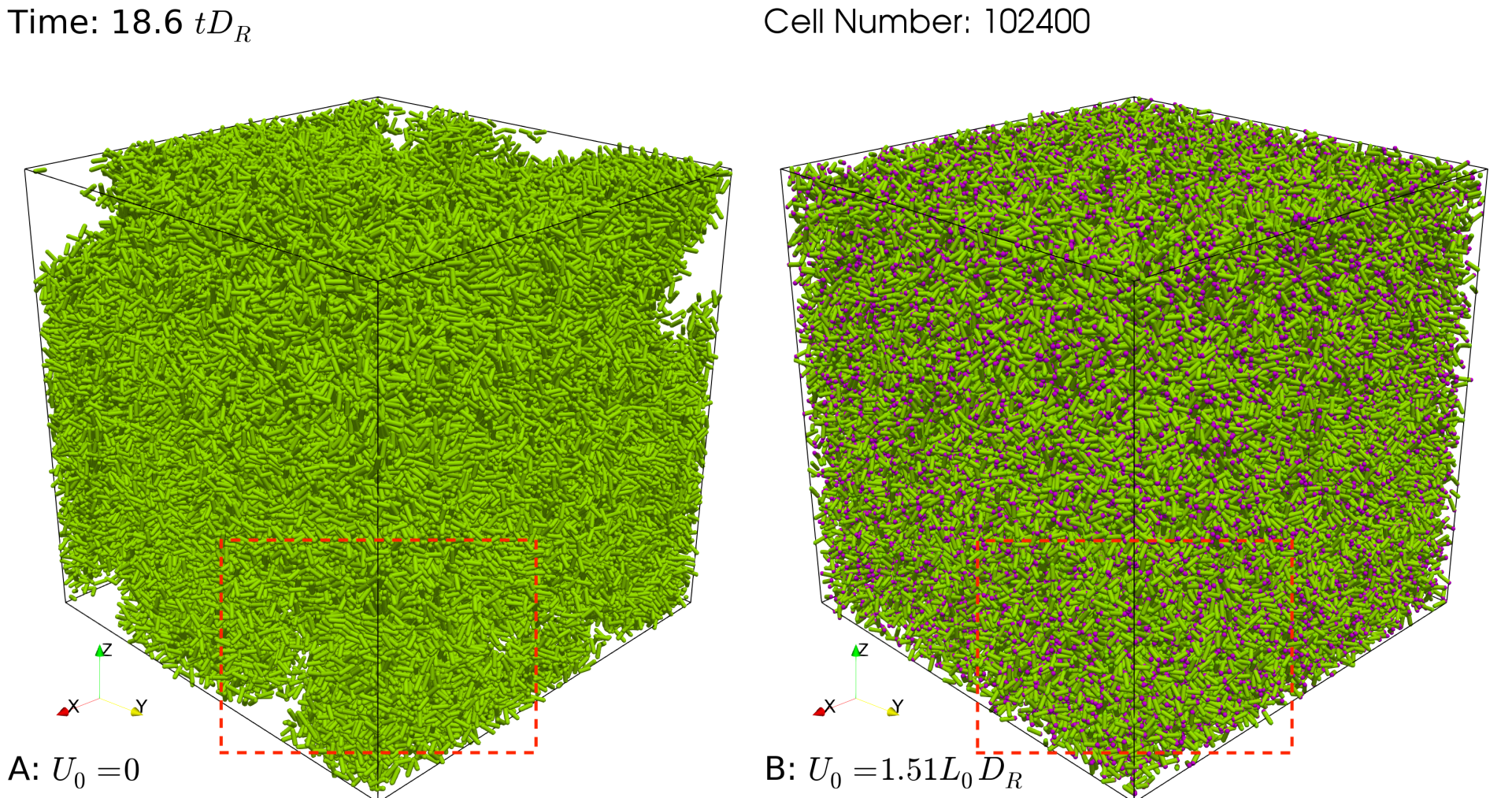}	
	\includegraphics[width=\linewidth]{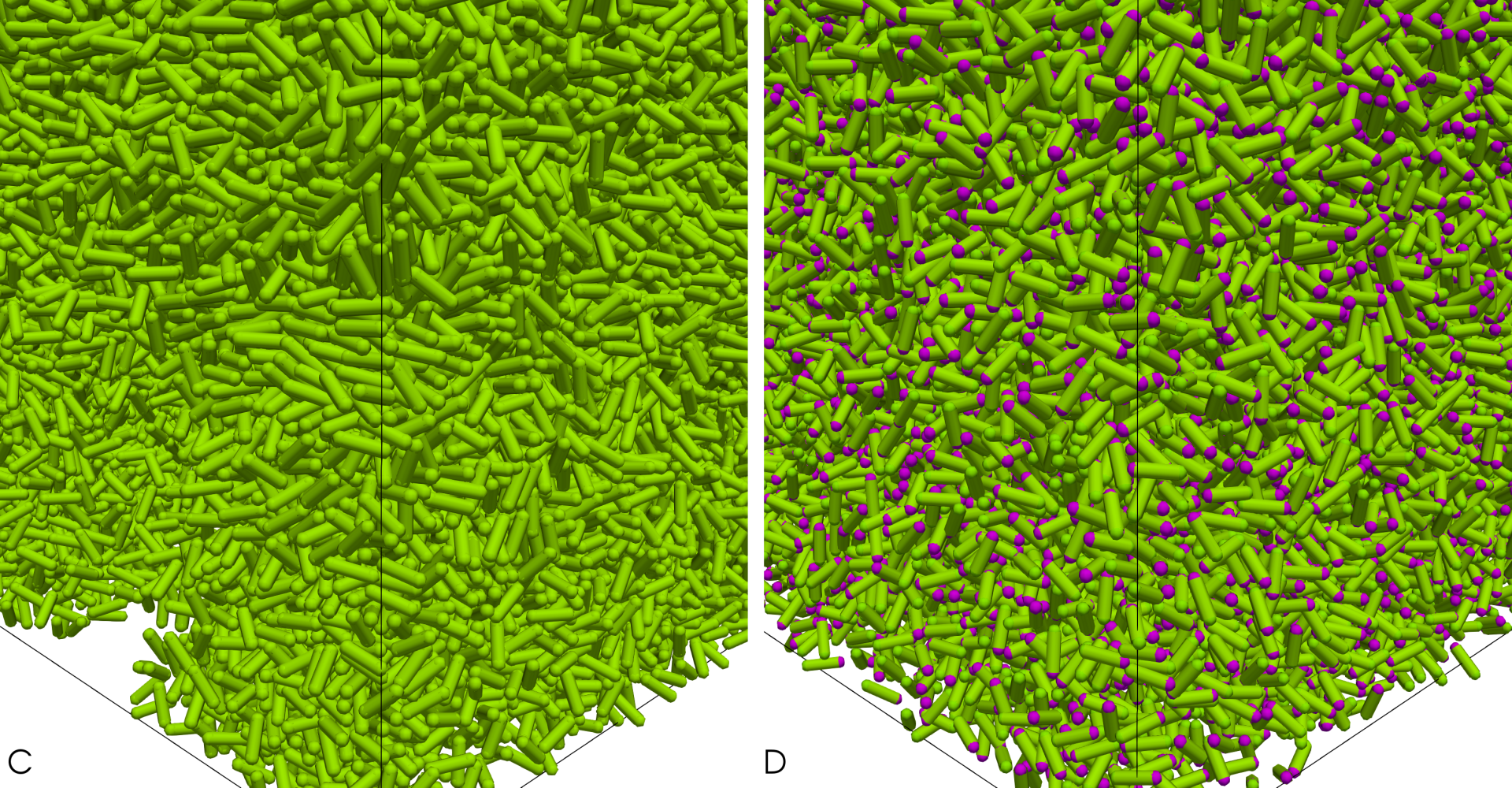}	
	\caption{The snapshot of dividing cells for A: $U_0=0$ and B: $U_0=1.51L_0D_R$ at time $tD_R=18.6$. 
	The purple dots mark the heads of the moving cells. 
	The red box marks the region shown in C and D.
	\label{fig:SnapCellDivision351}}
\end{figure}

We model biological cells as spherocylinders where the diameter $D$ remains constant but the length $L$ grows linearly in time.
All cells start to grow from a specified original length $L_0$ at $t=0$.
Once the length reaches the specified division length $L_D$, each cell splits into two shorter cells with equal length $L_0$. 
This division is assumed to occur instantaneously. 
As shown in Fig.~\ref{fig:Division}, we choose $L_D=2L_0+D$.
The new cells continue this growing-dividing cycle.
The number of cells in the simulation box therefore exponentially grows over time.
The division time  $\tau_{div}$ denotes the time one cell grows from $L_0$ to $L_D$, i.e., the time between two consecutive division events.

We use dimensional units: $D=\SI{1}{\micro\meter}$, $L_0= \SI{2.5}{\micro\meter}$, $L_D = \SI{6}{\micro\meter}$, and viscosity $\mu = \SI{0.001}{\pascal\cdot\second}$, close to the viscosity of water at room temperature. 
The Brownian motion is also computed as in the last section, where at room temperature $k_BT=\SI{0.00411}{\pico\newton\cdot\micro\meter}$.
All cells are assumed to divide at the same time. 
They are also assumed to swim in the direction $\bn$ with velocity $\bU_0=U_0\bn$ as the SPR model. 
All simulations start from 100 cells randomly and homogeneously distributed in a periodic cubic $100\times100\times100$ \si{\cubic\micro\meter} box. 

In this problem there are a variety of timescales, including the Brownian timescale $D_R^{-1}$, the swimming timescale $L_0/U_0$, the cell division timescale $\tau_{div}$, and the system relaxation timescale where the cell number density relaxes to a homogeneous distribution after each division.
A thorough investigation is beyond the scope of the current work, and we only report the results for a fast growing case where $\tau_{div}$ is longer than $D_R^{-1}$ but is much shorter than the density relaxation timescale.
We choose the rotational diffusion time $D_R^{-1}$ for cells with length $L_0$ as the unit of time. $D_R^{-1}=\SI{1.89}{\second}$ and we pick $\tau_{div}=\SI{3.5}{\second}$. 
 
\begin{figure}[htbp]
	\includegraphics[width=\linewidth]{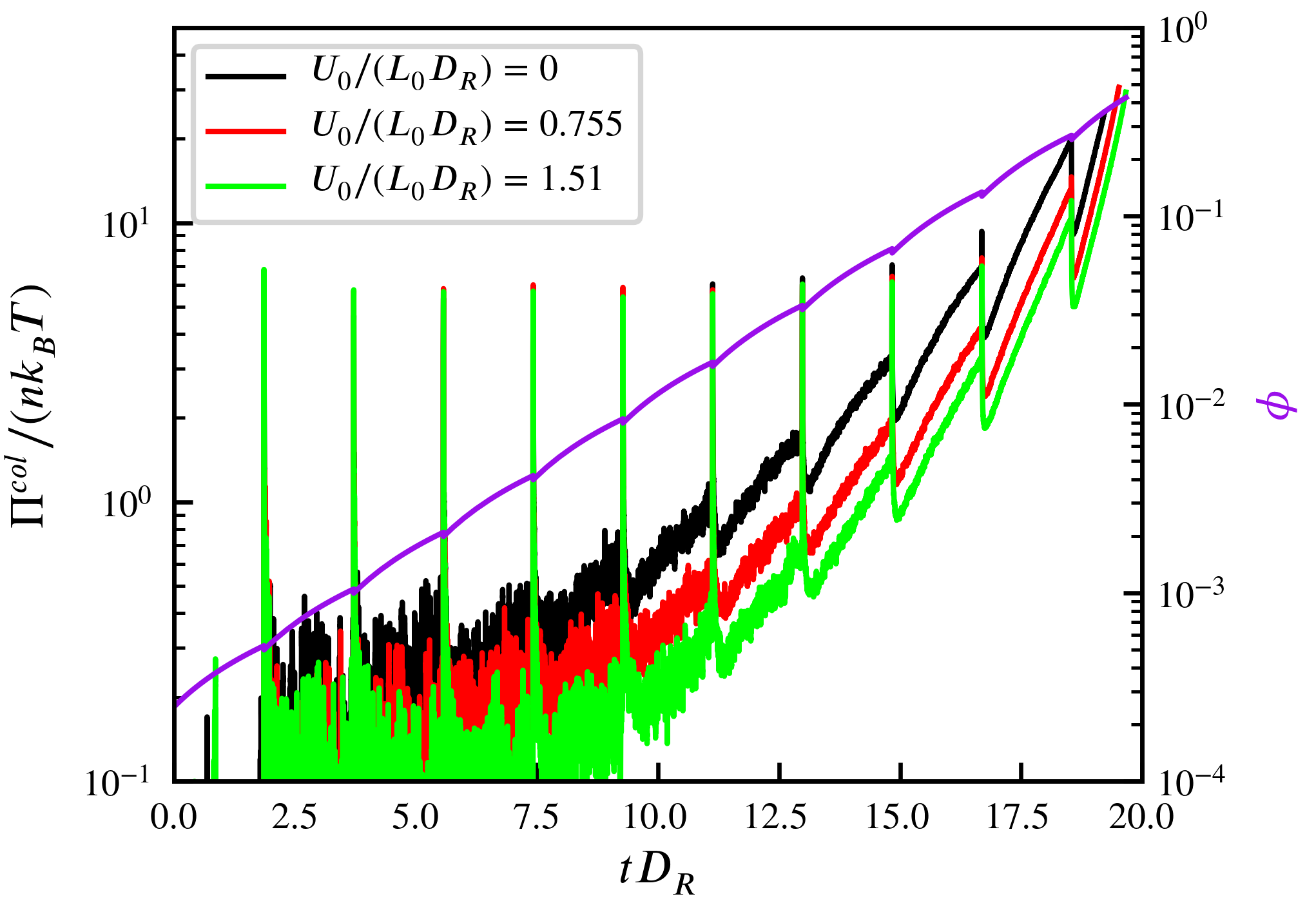}	
	\caption{The increase in collision pressure $\Pi^{col}$ for dividing cells with different self-propelled velocities $\bU_0$. 
	The timestep $\delta t = 5.3\times 10^{-5} D_R^{-1}$. 
	A moving average window of $100$ timesteps is applied to the measured $\Pi^{col}$ to filter the Brownian fluctuations.
	The purple line shows the exponential growth of the volume fraction $\phi$ over time. The tiny dips in $\phi$ at each collision event corresponds to the slight decrease in total cell volume as suggested in Fig.~\ref{fig:Division}.
	\label{fig:PiCellDivision}}
\end{figure}

The results are reported in dimensionless numbers in Fig.~\ref{fig:PiCellDivision}, where a moving average window of $100$ timesteps is applied to the measured $\Pi^{col}$ to filter the Brownian fluctuations. 
The measured collision pressure shows a peak, at almost the same height, at every division event before the volume fraction $\phi$ reaches $10\%$. 
This is because in dilute systems most collisions are contributed by those `newborn' pairs of cells with length $L_0$. 
This contribution is proportional to the total number of cells in the system, and therefore, when $\Pi^{col}$ is scaled by $nk_BT$ the total number is scaled out and the peaks are of almost the same height.

Another notable feature is that the systems with faster swimming velocity $U_0$ has lower collision pressure.
This is because the density relaxation time scale decreases with increasing $U_0$.
As shown in Fig.~\ref{fig:SnapCellDivision250}A and Fig.~\ref{fig:SnapCellDivision351}A, when $U_0=0$ the cells form local clusters because the division time $\tau_{div}$ is not sufficiently long for them to diffuse translationally.  
Such high density clusters increase the system collision pressure significantly.
While in Fig.~\ref{fig:SnapCellDivision250}B and Fig.~\ref{fig:SnapCellDivision351}B when $U_0=\SI{2}{\micro\meter\per\second}=1.51L_0D_R$, the system number density remains approximately homogeneous because of the swimming motion.

\section{Conclusion}
% summary
In this work we described a complete solution for computing the collision stress for moving rigid particle assemblies.
We first developed the general expression Eq.~(\ref{eq:colstressgeneral}) to compute the collision stress for each colliding pair of particles, based on the idea of volumetric integration of momentum transfer in that collision event. 
Equation~(\ref{eq:colstressgeneral}) is then demonstrated in Section~\ref{sec:paircolstress} to reproduce known expressions in various simplified cases.
This task can be completed by the LCP based collision resolution algorithm described in Section~\ref{sec:colalgo}. 
The idea is to utilize the geometric non-overlapping constraints and to remove the stiff pairwise repulsive potentials.
Our method is validated in Section~\ref{sec:benchsylinder} by measuring the system EOS for Brownian spherocylinders and finding accurate agreement with the work by \citet{bolhuis_tracing_1997}.
We further demonstrated briefly the applications of this method in Section~\ref{sec:application} for (i) self-propelled rods and (ii) growing-dividing cells. 
This new method allows us to measure mechanical properties in such soft active matter systems straightforwardly.

% extension to full Mobility
The method described in this work can be applied to various systems, as long as (i) the collision geometry for a pair of particles can be computed and (ii) the mobility matrix can be computed. 
We designed the method such that the mobility matrix $\bMcal$ appears only as an abstract matrix-vector multiplication operator. 
In this way $\bMcal$ can be computed with \emph{any} method without the necessity to explicitly construct the matrix, as long as the method keeps $\bMcal$ symmetric positive definite. 
In this paper we focused on the case where the many-body coupling in $\bMcal$ is ignored, i.e., $\bMcal$ is block-diagonal.
The same algorithm Eq.~(\ref{eq:LCPdef}) also works for the cases with full hydrodynamics.
For example, Rotne-Prager-Yamakawa tensor\cite{rotne_variational_1969}, Stokesian Dynamics\cite{wang_spectral_2016} and Boundary Integral method \cite{corona_boundary_2018} can all be used depending on the required accuracy for hydrodynamics for rigid particle suspensions.
We leave the analysis about the cases with full hydrodynamics to other forthcoming works.

% extension to other form of momentum exchange
Last, but not least, Eq.~(\ref{eq:colstressgeneral}) is applicable not only to the collision stress.
It is applicable to all cases where some form of momentum transfer happens from a point on one object to a point on another object.
Further, the impulse $\bJ$ does not have to be along the direction between the two points of momentum transfer. 
As long as the force $\bF_C$ and the geometry during the event can be computed, the stress follows Eq.~(\ref{eq:colstressgeneral}).
For example, in a microtubule network driven by motor proteins \cite{foster_connecting_2017}, the stress between microtubules generated by motor proteins can be computed with Eq.~(\ref{eq:colstressgeneral}) by replacing the force $\bF_C$ with the protein pushing or pulling force.
This paves the way to more fundamental understandings of the mechanical properties of such biological active networks.

\section{Acknowledgement}
MJS thanks the support from NSF Grants DMR-1420073 (NYUMRSEC), DMS-1463962, and DMS-1620331.

\appendix
\section{Geometry of spherocylinders.}
\label{sec:appsylinder}
Spherocylinders are cylinders of length $L$ and diameter $D$, capped with two hemispheres.
We define $\beta = L/D = L/(2R)$.
In the coordinate system where the spherocylinder is aligned with the $z$ axis, the integral $\bN$ and moment of inertia tensor $\bG_M$ are diagonalized:
\begin{align}
	\bN &= \rho\begin{bmatrix}
	N_\perp & 0 & 0 \\
	0 & N_\perp & 0 \\
	0 & 0 & N_\parallel \\
	\end{bmatrix},\\
		\bG_M &= \rho\begin{bmatrix}
	G_{M,\perp} & 0 & 0 \\
	0 & G_{M,\perp} & 0 \\
	0 & 0 & G_{M,\parallel} \\
	\end{bmatrix},
\end{align}
where
\begin{align}
N_{\perp}&=\frac{1}{30} (15 \beta+8) \pi  R^5,\\
N_{\parallel}&=\frac{1}{15} \left(10 \beta^3+20 \beta^2+15 \beta+4\right) \pi  R^5,\\
G_{M,\perp} &=\frac{1}{30} \left(20 \beta^3+40 \beta^2+45 \beta+16\right) \pi  R^5, \\
G_{M,\parallel} &=\frac{1}{15} (15 \beta+8) \pi  R^5.
\end{align}

\section{BBPGD}
\label{sec:appbbpgd}
This method can be summarized as the following algorithm:
	
\begin{algorithm}[H] % must use H, otherwise incompatible with revtex
	\caption{The Barzilai-Borwein Projected Gradient Descent method}
	\label{alg:BBPGD}
		\begin{algorithmic}
			\STATE Solve Eq.~\ref{eq:CQPdef} with initial guess $\bgamma_0$, residual tolerance $\epsilon_{tol}$, and $k_{max}$.
			\STATE $\bg_0=\bA\bgamma_0+\bb$.
			\IF{$\phi(\bgamma_0,\bg_0)<\epsilon$}
			  \STATE Solution is $\bgamma_0$.
			\ENDIF
			\STATE Simple gradient-descent step size $\alpha_0={\bg_0^T\bg_0}/{\bg_0^T\bA\bg_0}$.
			\FOR{ $k = 1 : k_{\max}$}
			  \STATE The descent step: $\bgamma_{k}= \bgamma_{k-1} - \alpha_{k-1} \bg_{k-1}$.
			  \STATE The projection step: $\bgamma_k=\Pi_{\bgamma\geq 0}\left[\bgamma_k\right]$.
			  \STATE Compute the gradient $\bg_k=\bA\bgamma_k+\bb$.
				\IF{$\varphi(\bgamma_k,\bg_k)\leq \epsilon_{tol}$}
				\STATE Stop iteration, solution is $\bgamma_k$.
				\ENDIF
			  \STATE $\bs_{k-1}=\bgamma_k-\bgamma_{k-1}$, $\by_{k-1}=\bg_k-\bg_{k-1}$.
			  \STATE $\alpha_k^{BB1}=\bs_{k-1}^T\bs_{k-1}/\bs_{k-1}^T\by_{k-1}$.
			\ENDFOR
		\end{algorithmic}
\end{algorithm}

In this algorithm $\alpha_k^{BB1}$ (next to the last line) is not the only choice. 
$\alpha_k^{BB2}=\bs_{k-1}^T\by_{k-1}/\by_{k-1}^T\by_{k-1}$ can also be used. 
% In fact, $\alpha_k^{BB1}$ and $\alpha_k^{BB2}$ can also be used alternatively for odd and even $k$-th steps.
We find that there is no significant difference in performance of different choices of $\alpha_k^{BB1}$ or $\alpha_k^{BB2}$ in solving our problems, and $\alpha_k^{BB1}$ is used for all results reported in this work.

\section{Collision between spherocylinders}
\label{sec:appcolsylinder}
\begin{figure}[h]
	\includegraphics[width=\linewidth]{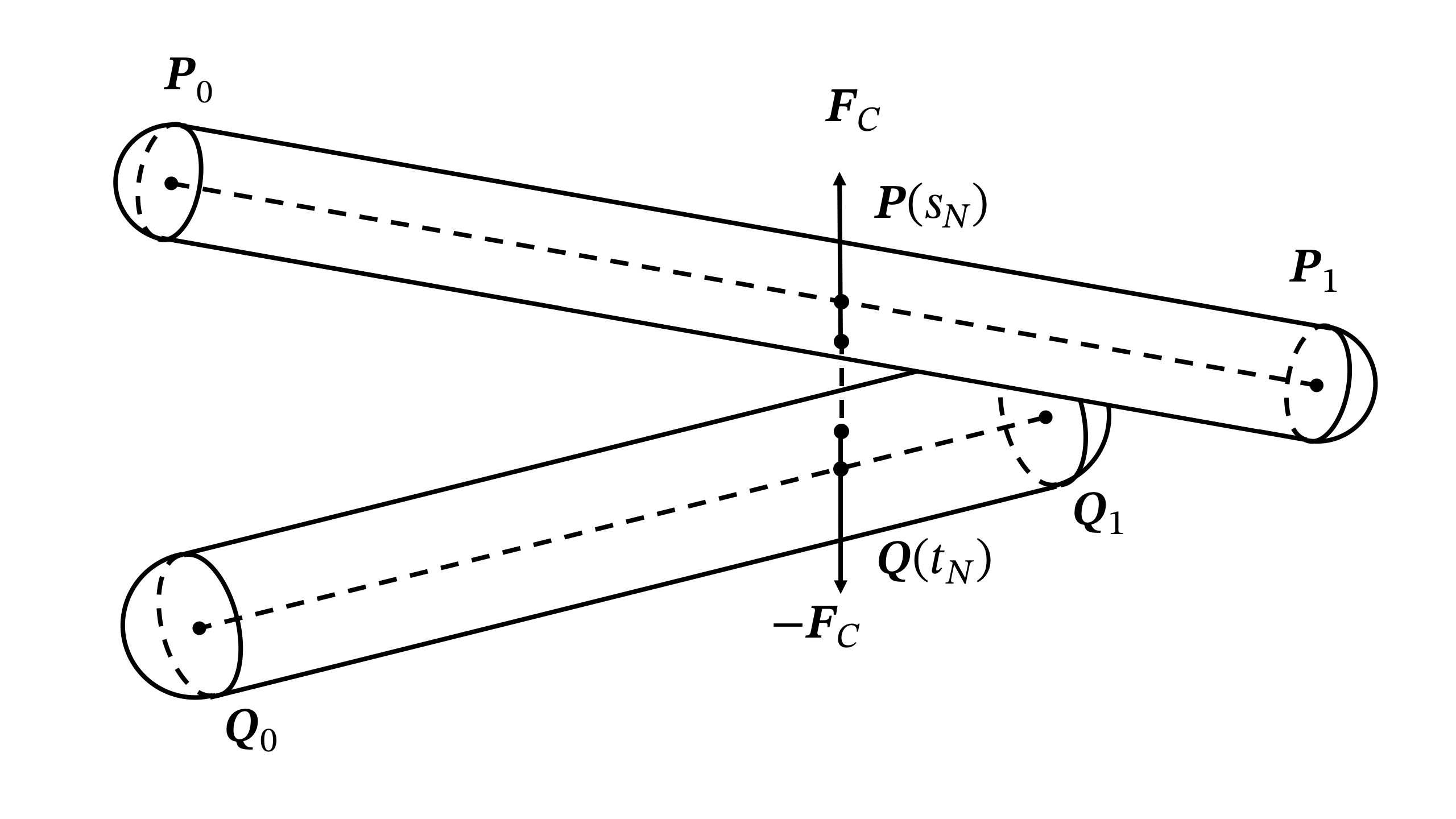}
	\caption{Collision geometry of two spherocylinders. \label{fig:colsegment}}
\end{figure}

% introduce how to compute the collision block for each colliding pair of sphero cylinders
This appendix describes how to find the minimum separation between a pair
of spherocylinders.
Geometrically, this task can be reduced to find the minimum distance between two line segments $\bP_0$, $\bP_1$, $\bQ_0$, $\bQ_1$ in 3D space,
 where $\bP_0,\bP_1$ (also $\bQ_0,\bQ_1$) are the two end points of the cylindrical section of one spherocylinder, as shown in Fig.~\ref{fig:colsegment}.

We parameterized the two spherocylinders with scalars $0<s,t<1$: $\bP(s)=(1-s)\bP_0+s\bP_1$ and
$\bQ(t)=(1-t)\bQ_0+t\bQ_1$.
Then the square distance between two
points on the segments is the quadratic function

\begin{align}
R(s,t)&=|\bP(s)-\bQ(t)|^2\\
 &=as^{2}-2bst+ct^{2}+2ds-2et+f\\
 &=\bp^{T}\bM\bp+2\bK^T\bp+f,
\label{eq:CGMobject}
\end{align}
where
\begin{align}
\bp^T &= \begin{bmatrix}
s & t	
\end{bmatrix},\\
\bM &= \begin{bmatrix}
	a & -b\\
	-b & c
\end{bmatrix},\\
\bK^T &= \begin{bmatrix}
d & -e	
\end{bmatrix},\\
a &=(\bP_1-\bP_0)\cdot(\bP_1-\bP_0),\\ 
b &=(\bP_1-\bP_0)\cdot(\bQ_1-\bQ_0),\\ 
c &=(\bQ_1-\bQ_0)\cdot(\bQ_1-\bQ_0),\\
d &=(\bP_1-\bP_0)\cdot(\bP_0-\bQ_0),\\ 
e &=(\bQ_1-\bQ_0)\cdot(\bP_0-\bQ_0),\\ 
f &=(\bP_0-\bQ_0)\cdot(\bP_0-\bQ_0).
\end{align}
$R(s,t)$ is a quadratic function to minimize on unit square $(s,t)\in[0,1]^2$.
Observe that 
\begin{align}\label{delta}
\mathrm{det}\bM &= ac-b^{2}\nonumber\\
 &= |(\bP_{1}-\bP_{0})\times(\bQ_{1}-\bQ_{0})|^{2}\geq0,
\end{align}
The minimization of $R(s,t)$ is straightforward, unless the two line segments are close to parallel, i.e., $\mathrm{det}\bM\to 0$. 
In this special case, numerical instabilities may occur due to the singularity of $\bM$. 
To handle all cases robustly, we follow the method described in the computational geometry library \texttt{Geometric Tools}\footnote{ David Eberly, Robust Computation of Distance Between Line Segments, https://www.geometrictools.com/}, where a constrained conjugate gradient approach is used. 
In our tests, this method computes the solution $s_N,t_N$ both efficiently and robustly. 

\begin{figure}[h]
	\includegraphics[width=\linewidth]{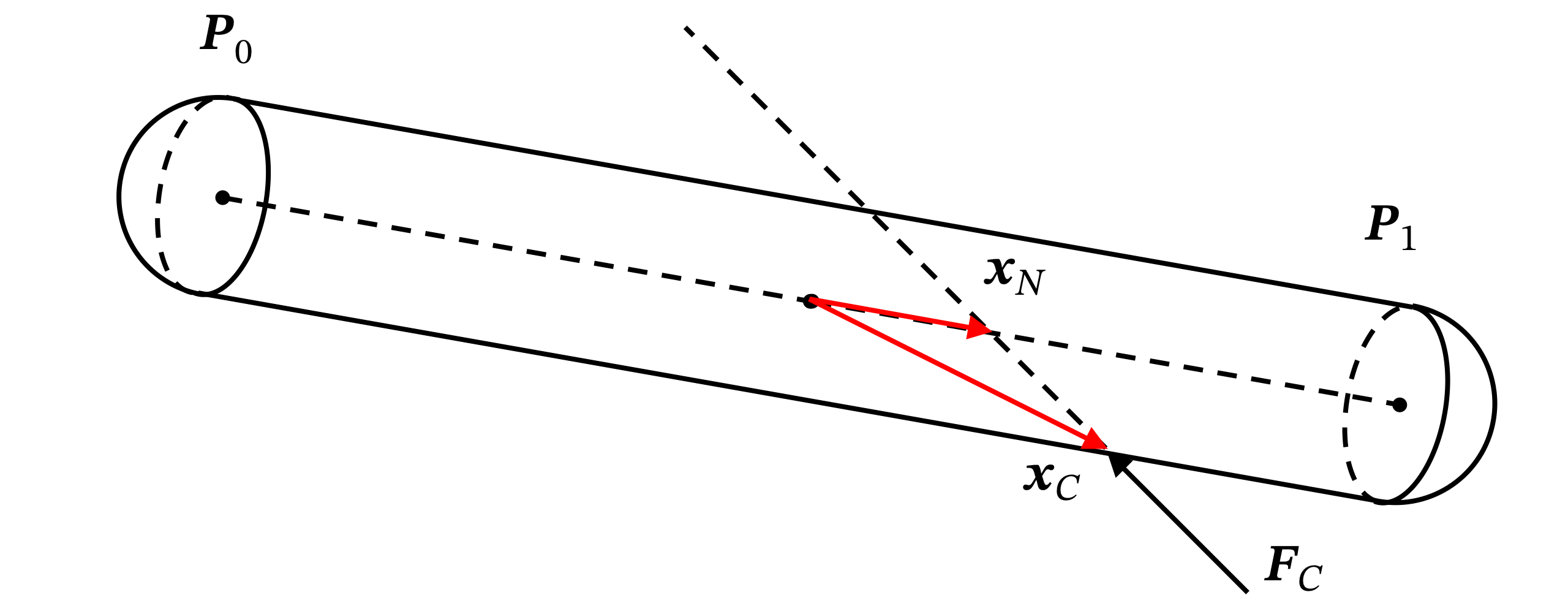}
	\caption{The relation between $\bx_N$ and $\bx_C$. \label{fig:xn_xc}}
\end{figure}

After we find $s_{N}$ and $t_{N}$ on each spherocylinder, we could easily compute the locations of minimal distance $\bP(s_{N})$ and $\bQ(t_{N})$.
The intersection points of vector $\bP(s_{N})-\bQ(t_{N})$ and surfaces of spherocylinders are the collision points. 
However, for the sake of convenience we do not need to find the exact collision points on surfaces.
When computing the stress tensor using Eq.~(\ref{eq:colstressgeneral}), only the torque relative to the center of mass $\bx_{C}\times\bF^{C}$ is necessary.
Geometrically it is straightforward to realize that $\bx_{C}\times\bF^{C}=\bx_{N}\times\bF^{C}$, as shown in Fig.~\ref{fig:xn_xc}.
Therefore there is no need to compute $\bx_C$.

\textbf{Reference}
%merlin.mbs aipnum4-1.bst 2010-07-25 4.21a (PWD, AO, DPC) hacked
%Control: key (0)
%Control: author (8) initials jnrlst
%Control: editor formatted (1) identically to author
%Control: production of article title (-1) disabled
%Control: page (0) single
%Control: year (1) truncated
%Control: production of eprint (0) enabled
%
	
% \bibliography{ref}

\begin{thebibliography}{38}%
	\makeatletter
	\providecommand \@ifxundefined [1]{%
	 \@ifx{#1\undefined}
	}%
	\providecommand \@ifnum [1]{%
	 \ifnum #1\expandafter \@firstoftwo
	 \else \expandafter \@secondoftwo
	 \fi
	}%
	\providecommand \@ifx [1]{%
	 \ifx #1\expandafter \@firstoftwo
	 \else \expandafter \@secondoftwo
	 \fi
	}%
	\providecommand \natexlab [1]{#1}%
	\providecommand \enquote  [1]{``#1''}%
	\providecommand \bibnamefont  [1]{#1}%
	\providecommand \bibfnamefont [1]{#1}%
	\providecommand \citenamefont [1]{#1}%
	\providecommand \href@noop [0]{\@secondoftwo}%
	\providecommand \href [0]{\begingroup \@sanitize@url \@href}%
	\providecommand \@href[1]{\@@startlink{#1}\@@href}%
	\providecommand \@@href[1]{\endgroup#1\@@endlink}%
	\providecommand \@sanitize@url [0]{\catcode `\\12\catcode `\$12\catcode
	  `\&12\catcode `\#12\catcode `\^12\catcode `\_12\catcode `\%12\relax}%
	\providecommand \@@startlink[1]{}%
	\providecommand \@@endlink[0]{}%
	\providecommand \url  [0]{\begingroup\@sanitize@url \@url }%
	\providecommand \@url [1]{\endgroup\@href {#1}{\urlprefix }}%
	\providecommand \urlprefix  [0]{URL }%
	\providecommand \Eprint [0]{\href }%
	\providecommand \doibase [0]{http://dx.doi.org/}%
	\providecommand \selectlanguage [0]{\@gobble}%
	\providecommand \bibinfo  [0]{\@secondoftwo}%
	\providecommand \bibfield  [0]{\@secondoftwo}%
	\providecommand \translation [1]{[#1]}%
	\providecommand \BibitemOpen [0]{}%
	\providecommand \bibitemStop [0]{}%
	\providecommand \bibitemNoStop [0]{.\EOS\space}%
	\providecommand \EOS [0]{\spacefactor3000\relax}%
	\providecommand \BibitemShut  [1]{\csname bibitem#1\endcsname}%
	\let\auto@bib@innerbib\@empty
	%</preamble>
	\bibitem [{\citenamefont {Bolhuis}\ and\ \citenamefont
	  {Frenkel}(1997)}]{bolhuis_tracing_1997}%
	  \BibitemOpen
	  \bibfield  {author} {\bibinfo {author} {\bibfnamefont {P.}~\bibnamefont
	  {Bolhuis}}\ and\ \bibinfo {author} {\bibfnamefont {D.}~\bibnamefont
	  {Frenkel}},\ }\href {\doibase 10.1063/1.473404} {\bibfield  {journal}
	  {\bibinfo  {journal} {The Journal of Chemical Physics}\ }\textbf {\bibinfo
	  {volume} {106}},\ \bibinfo {pages} {666} (\bibinfo {year}
	  {1997})}\BibitemShut {NoStop}%
	\bibitem [{\citenamefont {Takatori}, \citenamefont {Yan},\ and\ \citenamefont
	  {Brady}(2014)}]{TakatoriSwimPressureStress2014}%
	  \BibitemOpen
	  \bibfield  {author} {\bibinfo {author} {\bibfnamefont {S.~C.}\ \bibnamefont
	  {Takatori}}, \bibinfo {author} {\bibfnamefont {W.}~\bibnamefont {Yan}}, \
	  and\ \bibinfo {author} {\bibfnamefont {J.~F.}\ \bibnamefont {Brady}},\ }\href
	  {\doibase 10.1103/PhysRevLett.113.028103} {\bibfield  {journal} {\bibinfo
	  {journal} {Physical Review Letters}\ }\textbf {\bibinfo {volume} {113}},\
	  \bibinfo {pages} {028103} (\bibinfo {year} {2014})}\BibitemShut {NoStop}%
	\bibitem [{\citenamefont {Wang}\ and\ \citenamefont
	  {Brady}(2015)}]{wang_constant_2015}%
	  \BibitemOpen
	  \bibfield  {author} {\bibinfo {author} {\bibfnamefont {M.}~\bibnamefont
	  {Wang}}\ and\ \bibinfo {author} {\bibfnamefont {J.~F.}\ \bibnamefont
	  {Brady}},\ }\href {\doibase 10.1103/PhysRevLett.115.158301} {\bibfield
	  {journal} {\bibinfo  {journal} {Physical Review Letters}\ }\textbf {\bibinfo
	  {volume} {115}},\ \bibinfo {pages} {158301} (\bibinfo {year}
	  {2015})}\BibitemShut {NoStop}%
	\bibitem [{\citenamefont {Rebertus}\ and\ \citenamefont
	  {Sando}(1977)}]{rebertus_molecular_1977}%
	  \BibitemOpen
	  \bibfield  {author} {\bibinfo {author} {\bibfnamefont {D.~W.}\ \bibnamefont
	  {Rebertus}}\ and\ \bibinfo {author} {\bibfnamefont {K.~M.}\ \bibnamefont
	  {Sando}},\ }\href {\doibase 10.1063/1.435226} {\bibfield  {journal} {\bibinfo
	   {journal} {The Journal of Chemical Physics}\ }\textbf {\bibinfo {volume}
	  {67}},\ \bibinfo {pages} {2585} (\bibinfo {year} {1977})}\BibitemShut
	  {NoStop}%
	\bibitem [{\citenamefont {Snook}\ \emph {et~al.}(2014)\citenamefont {Snook},
	  \citenamefont {Davidson}, \citenamefont {Butler}, \citenamefont {Pouliquen},\
	  and\ \citenamefont {Guazzelli}}]{SnookNormalstressdifferences2014}%
	  \BibitemOpen
	  \bibfield  {author} {\bibinfo {author} {\bibfnamefont {B.}~\bibnamefont
	  {Snook}}, \bibinfo {author} {\bibfnamefont {L.~M.}\ \bibnamefont {Davidson}},
	  \bibinfo {author} {\bibfnamefont {J.~E.}\ \bibnamefont {Butler}}, \bibinfo
	  {author} {\bibfnamefont {O.}~\bibnamefont {Pouliquen}}, \ and\ \bibinfo
	  {author} {\bibfnamefont {E.}~\bibnamefont {Guazzelli}},\ }\href {\doibase
	  10.1017/jfm.2014.541} {\bibfield  {journal} {\bibinfo  {journal} {Journal of
	  Fluid Mechanics}\ }\textbf {\bibinfo {volume} {758}},\ \bibinfo {pages} {486}
	  (\bibinfo {year} {2014})}\BibitemShut {NoStop}%
	\bibitem [{\citenamefont {Campbell}\ and\ \citenamefont
	  {Gong}(1986)}]{Campbellstresstensortwodimensional1986}%
	  \BibitemOpen
	  \bibfield  {author} {\bibinfo {author} {\bibfnamefont {C.~S.}\ \bibnamefont
	  {Campbell}}\ and\ \bibinfo {author} {\bibfnamefont {A.}~\bibnamefont
	  {Gong}},\ }\href {\doibase 10.1017/S0022112086002495} {\bibfield  {journal}
	  {\bibinfo  {journal} {Journal of Fluid Mechanics}\ }\textbf {\bibinfo
	  {volume} {164}},\ \bibinfo {pages} {107} (\bibinfo {year}
	  {1986})}\BibitemShut {NoStop}%
	\bibitem [{\citenamefont {Campbell}(1989)}]{Campbellstresstensorsimple1989}%
	  \BibitemOpen
	  \bibfield  {author} {\bibinfo {author} {\bibfnamefont {C.~S.}\ \bibnamefont
	  {Campbell}},\ }\href {\doibase 10.1017/S0022112089001540} {\bibfield
	  {journal} {\bibinfo  {journal} {Journal of Fluid Mechanics}\ }\textbf
	  {\bibinfo {volume} {203}},\ \bibinfo {pages} {449} (\bibinfo {year}
	  {1989})}\BibitemShut {NoStop}%
	\bibitem [{\citenamefont {Tao}\ \emph {et~al.}(2005)\citenamefont {Tao},
	  \citenamefont {{den Otter}}, \citenamefont {Padding}, \citenamefont {Dhont},\
	  and\ \citenamefont {Briels}}]{tao_brownian_2005-1}%
	  \BibitemOpen
	  \bibfield  {author} {\bibinfo {author} {\bibfnamefont {Y.-G.}\ \bibnamefont
	  {Tao}}, \bibinfo {author} {\bibfnamefont {W.~K.}\ \bibnamefont {{den
	  Otter}}}, \bibinfo {author} {\bibfnamefont {J.~T.}\ \bibnamefont {Padding}},
	  \bibinfo {author} {\bibfnamefont {J.~K.~G.}\ \bibnamefont {Dhont}}, \ and\
	  \bibinfo {author} {\bibfnamefont {W.~J.}\ \bibnamefont {Briels}},\ }\href
	  {\doibase 10.1063/1.1940031} {\bibfield  {journal} {\bibinfo  {journal} {The
	  Journal of Chemical Physics}\ }\textbf {\bibinfo {volume} {122}},\ \bibinfo
	  {pages} {244903} (\bibinfo {year} {2005})}\BibitemShut {NoStop}%
	\bibitem [{\citenamefont {Maury}(2006)}]{Maurytimesteppingschemeinelastic2006}%
	  \BibitemOpen
	  \bibfield  {author} {\bibinfo {author} {\bibfnamefont {B.}~\bibnamefont
	  {Maury}},\ }\href {\doibase 10.1007/s00211-005-0666-6} {\bibfield  {journal}
	  {\bibinfo  {journal} {Numerische Mathematik}\ }\textbf {\bibinfo {volume}
	  {102}},\ \bibinfo {pages} {649} (\bibinfo {year} {2006})}\BibitemShut
	  {NoStop}%
	\bibitem [{\citenamefont {Tasora}, \citenamefont {Negrut},\ and\ \citenamefont
	  {Anitescu}(2008)}]{tasora_large-scale_2008}%
	  \BibitemOpen
	  \bibfield  {author} {\bibinfo {author} {\bibfnamefont {A.}~\bibnamefont
	  {Tasora}}, \bibinfo {author} {\bibfnamefont {D.}~\bibnamefont {Negrut}}, \
	  and\ \bibinfo {author} {\bibfnamefont {M.}~\bibnamefont {Anitescu}},\ }\href
	  {\doibase 10.1243/14644193JMBD154} {\bibfield  {journal} {\bibinfo  {journal}
	  {Proceedings of the Institution of Mechanical Engineers, Part K: Journal of
	  Multi-body Dynamics}\ }\textbf {\bibinfo {volume} {222}},\ \bibinfo {pages}
	  {315} (\bibinfo {year} {2008})}\BibitemShut {NoStop}%
	\bibitem [{\citenamefont {Tasora}\ and\ \citenamefont
	  {Anitescu}(2011)}]{tasora_matrix-free_2011}%
	  \BibitemOpen
	  \bibfield  {author} {\bibinfo {author} {\bibfnamefont {A.}~\bibnamefont
	  {Tasora}}\ and\ \bibinfo {author} {\bibfnamefont {M.}~\bibnamefont
	  {Anitescu}},\ }\href {\doibase 10.1016/j.cma.2010.06.030} {\bibfield
	  {journal} {\bibinfo  {journal} {Computer Methods in Applied Mechanics and
	  Engineering}\ }\textbf {\bibinfo {volume} {200}},\ \bibinfo {pages} {439}
	  (\bibinfo {year} {2011})}\BibitemShut {NoStop}%
	\bibitem [{\citenamefont {Foss}\ and\ \citenamefont
	  {Brady}(2000)}]{foss_brownian_2000}%
	  \BibitemOpen
	  \bibfield  {author} {\bibinfo {author} {\bibfnamefont {D.~R.}\ \bibnamefont
	  {Foss}}\ and\ \bibinfo {author} {\bibfnamefont {J.~F.}\ \bibnamefont
	  {Brady}},\ }\href {\doibase 10.1122/1.551104} {\bibfield  {journal} {\bibinfo
	   {journal} {Journal of Rheology}\ }\textbf {\bibinfo {volume} {44}},\
	  \bibinfo {pages} {629} (\bibinfo {year} {2000})}\BibitemShut {NoStop}%
	\bibitem [{\citenamefont {Kim}\ and\ \citenamefont
	  {Karrila}(2005)}]{Kim_Karrila_2005}%
	  \BibitemOpen
	  \bibfield  {author} {\bibinfo {author} {\bibfnamefont {S.}~\bibnamefont
	  {Kim}}\ and\ \bibinfo {author} {\bibfnamefont {S.~J.}\ \bibnamefont
	  {Karrila}},\ }\href@noop {} {\emph {\bibinfo {title} {Microhydrodynamics:
	  Principles and Selected Applications}}}\ (\bibinfo  {publisher} {Courier
	  Corporation},\ \bibinfo {year} {2005})\BibitemShut {NoStop}%
	\bibitem [{\citenamefont {Wang}\ and\ \citenamefont
	  {Brady}(2016)}]{wang_spectral_2016}%
	  \BibitemOpen
	  \bibfield  {author} {\bibinfo {author} {\bibfnamefont {M.}~\bibnamefont
	  {Wang}}\ and\ \bibinfo {author} {\bibfnamefont {J.~F.}\ \bibnamefont
	  {Brady}},\ }\href {\doibase 10.1016/j.jcp.2015.11.042} {\bibfield  {journal}
	  {\bibinfo  {journal} {Journal of Computational Physics}\ }\textbf {\bibinfo
	  {volume} {306}},\ \bibinfo {pages} {443} (\bibinfo {year}
	  {2016})}\BibitemShut {NoStop}%
	\bibitem [{\citenamefont {Corona}\ \emph {et~al.}(2017)\citenamefont {Corona},
	  \citenamefont {Greengard}, \citenamefont {Rachh},\ and\ \citenamefont
	  {Veerapaneni}}]{corona_integral_2017}%
	  \BibitemOpen
	  \bibfield  {author} {\bibinfo {author} {\bibfnamefont {E.}~\bibnamefont
	  {Corona}}, \bibinfo {author} {\bibfnamefont {L.}~\bibnamefont {Greengard}},
	  \bibinfo {author} {\bibfnamefont {M.}~\bibnamefont {Rachh}}, \ and\ \bibinfo
	  {author} {\bibfnamefont {S.}~\bibnamefont {Veerapaneni}},\ }\href {\doibase
	  10.1016/j.jcp.2016.12.018} {\bibfield  {journal} {\bibinfo  {journal}
	  {Journal of Computational Physics}\ }\textbf {\bibinfo {volume} {332}},\
	  \bibinfo {pages} {504} (\bibinfo {year} {2017})}\BibitemShut {NoStop}%
	\bibitem [{\citenamefont {Corona}\ and\ \citenamefont
	  {Veerapaneni}(2018)}]{corona_boundary_2018}%
	  \BibitemOpen
	  \bibfield  {author} {\bibinfo {author} {\bibfnamefont {E.}~\bibnamefont
	  {Corona}}\ and\ \bibinfo {author} {\bibfnamefont {S.}~\bibnamefont
	  {Veerapaneni}},\ }\href {\doibase 10.1016/j.jcp.2018.02.017} {\bibfield
	  {journal} {\bibinfo  {journal} {Journal of Computational Physics}\ }\textbf
	  {\bibinfo {volume} {362}},\ \bibinfo {pages} {327} (\bibinfo {year}
	  {2018})}\BibitemShut {NoStop}%
	\bibitem [{\citenamefont {Tornberg}\ and\ \citenamefont
	  {Gustavsson}(2006)}]{tornberg_numerical_2006}%
	  \BibitemOpen
	  \bibfield  {author} {\bibinfo {author} {\bibfnamefont {A.-K.}\ \bibnamefont
	  {Tornberg}}\ and\ \bibinfo {author} {\bibfnamefont {K.}~\bibnamefont
	  {Gustavsson}},\ }\href {\doibase 10.1016/j.jcp.2005.10.028} {\bibfield
	  {journal} {\bibinfo  {journal} {Journal of Computational Physics}\ }\textbf
	  {\bibinfo {volume} {215}},\ \bibinfo {pages} {172} (\bibinfo {year}
	  {2006})}\BibitemShut {NoStop}%
	\bibitem [{\citenamefont {Gustavsson}\ and\ \citenamefont
	  {Tornberg}(2009)}]{gustavsson_gravity_2009}%
	  \BibitemOpen
	  \bibfield  {author} {\bibinfo {author} {\bibfnamefont {K.}~\bibnamefont
	  {Gustavsson}}\ and\ \bibinfo {author} {\bibfnamefont {A.-K.}\ \bibnamefont
	  {Tornberg}},\ }\href {\doibase 10.1063/1.3273091} {\bibfield  {journal}
	  {\bibinfo  {journal} {Physics of Fluids (1994-present)}\ }\textbf {\bibinfo
	  {volume} {21}},\ \bibinfo {pages} {123301} (\bibinfo {year}
	  {2009})}\BibitemShut {NoStop}%
	\bibitem [{\citenamefont {Fang}(1984)}]{fang_linearization_1984}%
	  \BibitemOpen
	  \bibfield  {author} {\bibinfo {author} {\bibfnamefont {S.}~\bibnamefont
	  {Fang}},\ }\href {\doibase 10.1109/TAC.1984.1103393} {\bibfield  {journal}
	  {\bibinfo  {journal} {IEEE Transactions on Automatic Control}\ }\textbf
	  {\bibinfo {volume} {29}},\ \bibinfo {pages} {930} (\bibinfo {year}
	  {1984})}\BibitemShut {NoStop}%
	\bibitem [{\citenamefont {Niebe}\ and\ \citenamefont
	  {Erleben}(2015)}]{niebe_numerical_2015}%
	  \BibitemOpen
	  \bibfield  {author} {\bibinfo {author} {\bibfnamefont {S.}~\bibnamefont
	  {Niebe}}\ and\ \bibinfo {author} {\bibfnamefont {K.}~\bibnamefont
	  {Erleben}},\ }\href@noop {} {\emph {\bibinfo {title} {Numerical Methods for
	  Linear Complementarity Problems in Physics-Based Animation}}}\ (\bibinfo
	  {publisher} {{Morgan \& Claypool Publishers}},\ \bibinfo {address} {San
	  Rafael, California},\ \bibinfo {year} {2015})\ \bibinfo {note} {oCLC:
	  904469157}\BibitemShut {NoStop}%
	\bibitem [{\citenamefont {Mazhar}\ \emph {et~al.}(2015)\citenamefont {Mazhar},
	  \citenamefont {Heyn}, \citenamefont {Negrut},\ and\ \citenamefont
	  {Tasora}}]{mazhar_using_2015}%
	  \BibitemOpen
	  \bibfield  {author} {\bibinfo {author} {\bibfnamefont {H.}~\bibnamefont
	  {Mazhar}}, \bibinfo {author} {\bibfnamefont {T.}~\bibnamefont {Heyn}},
	  \bibinfo {author} {\bibfnamefont {D.}~\bibnamefont {Negrut}}, \ and\ \bibinfo
	  {author} {\bibfnamefont {A.}~\bibnamefont {Tasora}},\ }\href {\doibase
	  10.1145/2735627} {\bibfield  {journal} {\bibinfo  {journal} {ACM Trans.
	  Graph.}\ }\textbf {\bibinfo {volume} {34}},\ \bibinfo {pages} {32:1}
	  (\bibinfo {year} {2015})}\BibitemShut {NoStop}%
	\bibitem [{\citenamefont {Dai}\ and\ \citenamefont
	  {Fletcher}(2005)}]{dai_projected_2005}%
	  \BibitemOpen
	  \bibfield  {author} {\bibinfo {author} {\bibfnamefont {Y.-H.}\ \bibnamefont
	  {Dai}}\ and\ \bibinfo {author} {\bibfnamefont {R.}~\bibnamefont {Fletcher}},\
	  }\href {\doibase 10.1007/s00211-004-0569-y} {\bibfield  {journal} {\bibinfo
	  {journal} {Numerische Mathematik}\ }\textbf {\bibinfo {volume} {100}},\
	  \bibinfo {pages} {21} (\bibinfo {year} {2005})}\BibitemShut {NoStop}%
	\bibitem [{\citenamefont {L\"owen}(1994)}]{lowen_brownian_1994}%
	  \BibitemOpen
	  \bibfield  {author} {\bibinfo {author} {\bibfnamefont {H.}~\bibnamefont
	  {L\"owen}},\ }\href {\doibase 10.1103/PhysRevE.50.1232} {\bibfield  {journal}
	  {\bibinfo  {journal} {Physical Review E}\ }\textbf {\bibinfo {volume} {50}},\
	  \bibinfo {pages} {1232} (\bibinfo {year} {1994})}\BibitemShut {NoStop}%
	\bibitem [{\citenamefont {Delong}, \citenamefont {Usabiaga},\ and\
	  \citenamefont {Donev}(2015)}]{delong_brownian_2015}%
	  \BibitemOpen
	  \bibfield  {author} {\bibinfo {author} {\bibfnamefont {S.}~\bibnamefont
	  {Delong}}, \bibinfo {author} {\bibfnamefont {F.~B.}\ \bibnamefont
	  {Usabiaga}}, \ and\ \bibinfo {author} {\bibfnamefont {A.}~\bibnamefont
	  {Donev}},\ }\href {\doibase 10.1063/1.4932062} {\bibfield  {journal}
	  {\bibinfo  {journal} {The Journal of Chemical Physics}\ }\textbf {\bibinfo
	  {volume} {143}},\ \bibinfo {pages} {144107} (\bibinfo {year}
	  {2015})}\BibitemShut {NoStop}%
	\bibitem [{\citenamefont {Baskaran}\ and\ \citenamefont
	  {Marchetti}(2008)}]{baskaran_enhanced_2008}%
	  \BibitemOpen
	  \bibfield  {author} {\bibinfo {author} {\bibfnamefont {A.}~\bibnamefont
	  {Baskaran}}\ and\ \bibinfo {author} {\bibfnamefont {M.~C.}\ \bibnamefont
	  {Marchetti}},\ }\href {\doibase 10.1103/PhysRevLett.101.268101} {\bibfield
	  {journal} {\bibinfo  {journal} {Physical Review Letters}\ }\textbf {\bibinfo
	  {volume} {101}},\ \bibinfo {pages} {268101} (\bibinfo {year}
	  {2008})}\BibitemShut {NoStop}%
	\bibitem [{\citenamefont {Ginelli}\ \emph {et~al.}(2010)\citenamefont
	  {Ginelli}, \citenamefont {Peruani}, \citenamefont {B\"ar},\ and\
	  \citenamefont {Chat\'e}}]{ginelli_large-scale_2010}%
	  \BibitemOpen
	  \bibfield  {author} {\bibinfo {author} {\bibfnamefont {F.}~\bibnamefont
	  {Ginelli}}, \bibinfo {author} {\bibfnamefont {F.}~\bibnamefont {Peruani}},
	  \bibinfo {author} {\bibfnamefont {M.}~\bibnamefont {B\"ar}}, \ and\ \bibinfo
	  {author} {\bibfnamefont {H.}~\bibnamefont {Chat\'e}},\ }\href {\doibase
	  10.1103/PhysRevLett.104.184502} {\bibfield  {journal} {\bibinfo  {journal}
	  {Physical Review Letters}\ }\textbf {\bibinfo {volume} {104}},\ \bibinfo
	  {pages} {184502} (\bibinfo {year} {2010})}\BibitemShut {NoStop}%
	\bibitem [{\citenamefont {{Orozco-Fuentes}}\ and\ \citenamefont
	  {Boyer}(2013)}]{orozco-fuentes_order_2013}%
	  \BibitemOpen
	  \bibfield  {author} {\bibinfo {author} {\bibfnamefont {S.}~\bibnamefont
	  {{Orozco-Fuentes}}}\ and\ \bibinfo {author} {\bibfnamefont {D.}~\bibnamefont
	  {Boyer}},\ }\href {\doibase 10.1103/PhysRevE.88.012715} {\bibfield  {journal}
	  {\bibinfo  {journal} {Physical Review E}\ }\textbf {\bibinfo {volume} {88}}
	  (\bibinfo {year} {2013}),\ 10.1103/PhysRevE.88.012715}\BibitemShut {NoStop}%
	\bibitem [{\citenamefont {Kuan}\ \emph {et~al.}(2015)\citenamefont {Kuan},
	  \citenamefont {Blackwell}, \citenamefont {Hough}, \citenamefont {Glaser},\
	  and\ \citenamefont {Betterton}}]{kuan_hysteresis_2015-1}%
	  \BibitemOpen
	  \bibfield  {author} {\bibinfo {author} {\bibfnamefont {H.-S.}\ \bibnamefont
	  {Kuan}}, \bibinfo {author} {\bibfnamefont {R.}~\bibnamefont {Blackwell}},
	  \bibinfo {author} {\bibfnamefont {L.~E.}\ \bibnamefont {Hough}}, \bibinfo
	  {author} {\bibfnamefont {M.~A.}\ \bibnamefont {Glaser}}, \ and\ \bibinfo
	  {author} {\bibfnamefont {M.~D.}\ \bibnamefont {Betterton}},\ }\href {\doibase
	  10.1103/PhysRevE.92.060501} {\bibfield  {journal} {\bibinfo  {journal}
	  {Physical Review E}\ }\textbf {\bibinfo {volume} {92}},\ \bibinfo {pages}
	  {060501} (\bibinfo {year} {2015})}\BibitemShut {NoStop}%
	\bibitem [{\citenamefont {Weitz}, \citenamefont {Deutsch},\ and\ \citenamefont
	  {Peruani}(2015)}]{weitz_self-propelled_2015-1}%
	  \BibitemOpen
	  \bibfield  {author} {\bibinfo {author} {\bibfnamefont {S.}~\bibnamefont
	  {Weitz}}, \bibinfo {author} {\bibfnamefont {A.}~\bibnamefont {Deutsch}}, \
	  and\ \bibinfo {author} {\bibfnamefont {F.}~\bibnamefont {Peruani}},\ }\href
	  {\doibase 10.1103/PhysRevE.92.012322} {\bibfield  {journal} {\bibinfo
	  {journal} {Physical Review E}\ }\textbf {\bibinfo {volume} {92}},\ \bibinfo
	  {pages} {012322} (\bibinfo {year} {2015})}\BibitemShut {NoStop}%
	\bibitem [{\citenamefont {Peruani}(2016)}]{peruani_active_2016}%
	  \BibitemOpen
	  \bibfield  {author} {\bibinfo {author} {\bibfnamefont {F.}~\bibnamefont
	  {Peruani}},\ }\href {\doibase 10.1140/epjst/e2016-60062-0} {\bibfield
	  {journal} {\bibinfo  {journal} {The European Physical Journal Special
	  Topics}\ }\textbf {\bibinfo {volume} {225}},\ \bibinfo {pages} {2301}
	  (\bibinfo {year} {2016})}\BibitemShut {NoStop}%
	\bibitem [{\citenamefont {Gro\ss{}mann}, \citenamefont {Peruani},\ and\
	  \citenamefont {B\"ar}(2016)}]{grosmann_mesoscale_2016}%
	  \BibitemOpen
	  \bibfield  {author} {\bibinfo {author} {\bibfnamefont {R.}~\bibnamefont
	  {Gro\ss{}mann}}, \bibinfo {author} {\bibfnamefont {F.}~\bibnamefont
	  {Peruani}}, \ and\ \bibinfo {author} {\bibfnamefont {M.}~\bibnamefont
	  {B\"ar}},\ }\href {\doibase 10.1103/PhysRevE.94.050602} {\bibfield  {journal}
	  {\bibinfo  {journal} {Physical Review E}\ }\textbf {\bibinfo {volume} {94}},\
	  \bibinfo {pages} {050602} (\bibinfo {year} {2016})}\BibitemShut {NoStop}%
	\bibitem [{\citenamefont {Vroege}\ and\ \citenamefont
	  {Lekkerkerker}(1992)}]{vroege_phase_1992}%
	  \BibitemOpen
	  \bibfield  {author} {\bibinfo {author} {\bibfnamefont {G.~J.}\ \bibnamefont
	  {Vroege}}\ and\ \bibinfo {author} {\bibfnamefont {H.~N.~W.}\ \bibnamefont
	  {Lekkerkerker}},\ }\href {\doibase 10.1088/0034-4885/55/8/003} {\bibfield
	  {journal} {\bibinfo  {journal} {Reports on Progress in Physics}\ }\textbf
	  {\bibinfo {volume} {55}},\ \bibinfo {pages} {1241} (\bibinfo {year}
	  {1992})}\BibitemShut {NoStop}%
	\bibitem [{\citenamefont {Onsager}(1949)}]{onsager_effects_1949}%
	  \BibitemOpen
	  \bibfield  {author} {\bibinfo {author} {\bibfnamefont {L.}~\bibnamefont
	  {Onsager}},\ }\href {\doibase 10.1111/j.1749-6632.1949.tb27296.x} {\bibfield
	  {journal} {\bibinfo  {journal} {Annals of the New York Academy of Sciences}\
	  }\textbf {\bibinfo {volume} {51}},\ \bibinfo {pages} {627} (\bibinfo {year}
	  {1949})}\BibitemShut {NoStop}%
	\bibitem [{\citenamefont {Graf}\ and\ \citenamefont
	  {L\"owen}(1999)}]{graf_density_1999}%
	  \BibitemOpen
	  \bibfield  {author} {\bibinfo {author} {\bibfnamefont {H.}~\bibnamefont
	  {Graf}}\ and\ \bibinfo {author} {\bibfnamefont {H.}~\bibnamefont {L\"owen}},\
	  }\href {\doibase 10.1088/0953-8984/11/6/008} {\bibfield  {journal} {\bibinfo
	  {journal} {Journal of Physics: Condensed Matter}\ }\textbf {\bibinfo {volume}
	  {11}},\ \bibinfo {pages} {1435} (\bibinfo {year} {1999})}\BibitemShut
	  {NoStop}%
	\bibitem [{\citenamefont {Kraikivski}, \citenamefont {Lipowsky},\ and\
	  \citenamefont {Kierfeld}(2006)}]{kraikivski_enhanced_2006}%
	  \BibitemOpen
	  \bibfield  {author} {\bibinfo {author} {\bibfnamefont {P.}~\bibnamefont
	  {Kraikivski}}, \bibinfo {author} {\bibfnamefont {R.}~\bibnamefont
	  {Lipowsky}}, \ and\ \bibinfo {author} {\bibfnamefont {J.}~\bibnamefont
	  {Kierfeld}},\ }\href {\doibase 10.1103/PhysRevLett.96.258103} {\bibfield
	  {journal} {\bibinfo  {journal} {Physical Review Letters}\ }\textbf {\bibinfo
	  {volume} {96}},\ \bibinfo {pages} {258103} (\bibinfo {year}
	  {2006})}\BibitemShut {NoStop}%
	\bibitem [{\citenamefont {Rotne}\ and\ \citenamefont
	  {Prager}(1969)}]{rotne_variational_1969}%
	  \BibitemOpen
	  \bibfield  {author} {\bibinfo {author} {\bibfnamefont {J.}~\bibnamefont
	  {Rotne}}\ and\ \bibinfo {author} {\bibfnamefont {S.}~\bibnamefont {Prager}},\
	  }\href {\doibase 10.1063/1.1670977} {\bibfield  {journal} {\bibinfo
	  {journal} {The Journal of Chemical Physics}\ }\textbf {\bibinfo {volume}
	  {50}},\ \bibinfo {pages} {4831} (\bibinfo {year} {1969})}\BibitemShut
	  {NoStop}%
	\bibitem [{\citenamefont {Foster}\ \emph {et~al.}(2017)\citenamefont {Foster},
	  \citenamefont {Yan}, \citenamefont {F\"urthauer}, \citenamefont {Shelley},\
	  and\ \citenamefont {Needleman}}]{foster_connecting_2017}%
	  \BibitemOpen
	  \bibfield  {author} {\bibinfo {author} {\bibfnamefont {P.~J.}\ \bibnamefont
	  {Foster}}, \bibinfo {author} {\bibfnamefont {W.}~\bibnamefont {Yan}},
	  \bibinfo {author} {\bibfnamefont {S.}~\bibnamefont {F\"urthauer}}, \bibinfo
	  {author} {\bibfnamefont {M.~J.}\ \bibnamefont {Shelley}}, \ and\ \bibinfo
	  {author} {\bibfnamefont {D.~J.}\ \bibnamefont {Needleman}},\ }\href {\doibase
	  10.1088/1367-2630/aa9320} {\bibfield  {journal} {\bibinfo  {journal} {New
	  Journal of Physics}\ }\textbf {\bibinfo {volume} {19}},\ \bibinfo {pages}
	  {125011} (\bibinfo {year} {2017})}\BibitemShut {NoStop}%
	\bibitem [{Note1()}]{Note1}%
	  \BibitemOpen
	  \bibinfo {note} {David Eberly, Robust Computation of Distance Between Line
	  Segments, https://www.geometrictools.com/}\BibitemShut {NoStop}%
	\end{thebibliography}

\end{document}